\documentclass[aps,prb,reprint,superscriptaddress]{revtex4-2}
\pdfoutput=1
\usepackage[english]{babel}
\usepackage[utf8x]{inputenc}
\usepackage[T1]{fontenc}
\usepackage{siunitx}
\usepackage{xcolor}
\usepackage{xspace}
\usepackage[version=4]{mhchem}

\usepackage{placeins}
\usepackage{braket}
\usepackage{longtable}
\usepackage{todonotes}
\usepackage{booktabs}
\usepackage{natbib}
\usepackage{bibentry}
\usepackage{amsmath}
\usepackage{graphicx}
\usepackage[nolist]{acronym}
\usepackage[colorlinks=true, allcolors=blue, breaklinks=true]{hyperref}

\usepackage{ragged2e}
\usepackage{makecell}

\DeclareSIUnit\plus{+}
\begin{document}

\title{Millisecond electron spin coherence time for erbium ions in silicon}
\author{Ian R. \surname{Berkman}}
\thanks{These two authors contributed equally to this work}
\affiliation{Centre of Excellence for Quantum Computation and Communication Technology, School of Physics, University of New South Wales, Sydney, NSW 2052, Australia}
\author{Alexey \surname{Lyasota}}
\thanks{These two authors contributed equally to this work}
\affiliation{Centre of Excellence for Quantum Computation and Communication Technology, School of Physics, University of New South Wales, Sydney, NSW 2052, Australia}
\author{Gabriele G. \surname{de Boo}}
\affiliation{Centre of Excellence for Quantum Computation and Communication Technology, School of Physics, University of New South Wales, Sydney, NSW 2052, Australia}
\author{John G. \surname{Bartholomew}}
\affiliation{Centre for Engineered Quantum Systems, School of Physics, The University of Sydney, Sydney, NSW 2006, Australia}
\affiliation{The University of Sydney Nano Institute, The University of Sydney, Sydney, NSW 2006, Australia}
\author{Shao Q. \surname{Lim}} 
\affiliation{Centre of Excellence for Quantum Computation and Communication Technology, School of Physics, University of Melbourne, Victoria 3010, Australia}
\author{Brett C. \surname{Johnson}}
\affiliation{Centre of Excellence for Quantum Computation and Communication Technology, School of Physics, University of Melbourne, Victoria 3010, Australia}
\affiliation{School of Science, RMIT University, Victoria 3001, Australia}
\author{Jeffrey C. \surname{McCallum}}
\affiliation{Centre of Excellence for Quantum Computation and Communication Technology, School of Physics, University of Melbourne, Victoria 3010, Australia}
\author{Bin-Bin \surname{Xu}}
\affiliation{Centre of Excellence for Quantum Computation and Communication Technology, School of Physics, University of New South Wales, Sydney, NSW 2052, Australia}
\author{Shouyi \surname{Xie}}
\affiliation{Centre of Excellence for Quantum Computation and Communication Technology, School of Physics, University of New South Wales, Sydney, NSW 2052, Australia}
\author{Nikolay V. \surname{Abrosimov}}
\affiliation{Leibniz-Institut f\"{u}r Kristallz\"{u}chtung, 12489 Berlin, Germany}
\author{Hans-Joachim \surname{Pohl}}
\affiliation{VITCON Projectconsult GmbH, 07745 Jena, Germany}
\author{Rose L. \surname{Ahlefeldt}}
\affiliation{Centre of Excellence for Quantum Computation and Communication Technology, Research School of Physics, Australian National University, Canberra, ACT 0200, Australia}
\author{Matthew J. \surname{Sellars}}
\affiliation{Centre of Excellence for Quantum Computation and Communication Technology, Research School of Physics, Australian National University, Canberra, ACT 0200, Australia}
\author{Chunming \surname{Yin}} 
\affiliation{Centre of Excellence for Quantum Computation and Communication Technology, School of Physics, University of New South Wales, Sydney, NSW 2052, Australia}
\affiliation{%
CAS Key Laboratory of Microscale Magnetic Resonance and School of Physical Sciences, University of Science and Technology of China, Hefei 230026, China
}
\author{Sven \surname{Rogge}}
\affiliation{Centre of Excellence for Quantum Computation and Communication Technology, School of Physics, University of New South Wales, Sydney, NSW 2052, Australia}

\newcommand{\erthree}{\ce{Er^{3+}}\xspace}
\newcommand{\er}{\ce{Er}\xspace}
\newcommand{\hethree}{\ce{^3He}\xspace}
\newcommand{\puresi}{\ce{^{28}Si}\xspace}
\newcommand{\sitwonine}{\ce{^{29}Si}\xspace}
\newcommand{\spinhalf}{spin-\num{1}/\num{2}\xspace}
\newcommand{\erinsi}{\ce{Er^{3+}}:\ce{Si}\xspace}
\newcommand{\tone}{$T_1$\xspace}
\newcommand{\ttworabi}{$T_{2,\text{Rabi}}$\xspace}
\newcommand{\ttwohahn}{$T_{2,\text{Hahn}}$\xspace}
\newcommand{\pitwox}{$\frac{\pi}{2}_x$\xspace}
\newcommand{\piy}{$\pi_y$\xspace}
\newcommand{\fifteenhalf}{\ce{^4I_{15/2}}\xspace}
\newcommand{\thirteenhalf}{\ce{^4I_{13/2}}\xspace}
\newcommand{\spinup}{$\ket{\uparrow}$\xspace}
\newcommand{\spindown}{$\ket{\downarrow}$\xspace}
\newcommand{\mszero}{$\Delta M_S = 0$\xspace}
\newcommand{\msone}{$\Delta M_S = 1$\xspace}
\newcommand{\Sitec}{Site E\xspace}
\newcommand{\Sited}{Site F\xspace}
\newcommand{\YSO}{\ce{Y_2SiO_5} \xspace}
\newcommand{\YVO}{\ce{YVO_4} \xspace}
\newcommand{\CaWO}{\ce{CaWO_4} \xspace}
\newcommand{\LiNbO}{\ce{LiNbO_3} \xspace}
\DeclareSIUnit\decibelm{dBm}
\DeclareSIUnit\belm{Bm}

\begin{abstract}
Spins in silicon that are accessible via a telecom-compatible optical transition are a versatile platform for quantum information processing that can leverage the well-established silicon nanofabrication industry. 
Key to these applications are long coherence times on the optical and spin transitions to provide a robust system for interfacing photonic and spin qubits. 
Here, we report telecom-compatible \erthree sites with long optical and electron spin coherence times, measured within a nuclear spin-free silicon crystal (\SI{<0.01}{\percent} \sitwonine) using optical detection. 
We investigate two sites and find \SI{0.1}{\giga\hertz} optical inhomogeneous linewidths and homogeneous linewidths below \SI{70}{\kilo\hertz} for both sites.
We measure the electron spin coherence time of both sites using optically detected magnetic resonance and observe Hahn echo decay constants of \SI{0.8}{\milli\second} and \SI{1.2}{\milli\second} at \SI{\sim11}{\milli\tesla}. 
These optical and spin properties of \erinsi are an important milestone towards using optically accessible spins in silicon for a broad range of quantum information processing applications.
\end{abstract}
\maketitle

\begin{acronym}
    \acro{RE}[RE]{rare earth}
	\acro{PL}[PL]{photoluminescence}
	\acro{PLE}[PLE]{photoluminescence excitation}
	\acro{ESR}[ESR]{electron spin resonance}
        \acro{ODMR}[ODMR]{optically detected magnetic resonance}
        \acro{RE}[RE]{rare-earth}
        \acro{MBE}[MBE]{molecular-beam epitaxy}
	\acro{SSPD}[SSPD]{superconducting single-photon detector}
	\acro{SOI}[SOI]{silicon-on-insulator}
	\acro{QIP}[QIP]{quantum information processing}
	\acro{VLSI}[VLSI]{very large-scale integration}
	\acro{CMOS}[CMOS]{complementary metal-oxide-semiconductor}
	\acro{SNR}[SNR]{signal-to-noise ratio}
	\acro{AOM}[AOM]{acousto-optical modulator}
	\acro{EOM}[EOM]{electro-optical modulator}
	\acro{HEMT}[HEMT]{high-electron-mobility transistor}
	\acro{FinFET}[FinFET]{fin field-effect transistor}
	\acro{FWHM}[FWHM]{full width at half maximum}
	\acro{RF}[RF]{radiofrequency}
\end{acronym}


\section*{Introduction}

\Ac{QIP} offers numerous protocols that outperform equivalent classical protocols in terms of computational speed and security~\cite{Nielsen2010,Shor1997,Grover1996,Wehner2018,Childs2010,DiVincenzo1995,Ekert1996,Feynman1982,Jordan2011,Montanaro2016,Alexeev2021}.
In order to enact these protocols, the ability to coherently manipulate and entangle multiple quantum states is required.
Electron spins in silicon are attractive qubits due to the fast control of the highly coherent spin states~\cite{Kane1998,Pla2012, Fricke2021,Philips2022,Burkard2021,Tyryshkin2012,Veldhorst2014,Takeda2022,Veldhorst2015,Xue2022,Yoneda2018,Bergeron2020} and the ability to use well-established semiconductor processing techniques to fabricate scalable nanometer architectures with high yield~\cite{ElKareh1995,Li2018,Jones2018,Zajac2016,Zwerver2022}.
Additionally, silicon systems that are capable of photon-mediated spin-spin coupling are promising candidates for inter-chip and chip-to-chip coupling in quantum computing~\cite{Vandersypen2017,Borjans2020,Yan2021,Harvey2022}, and for long-distance coupling in quantum networks~\cite{Kimble2008,Wehner2018,Borregaard2019,Zhu2020}.
For these spin-photon systems, it is favourable to minimise the photon losses between the spin qubits in order to achieve entanglement operations with high fidelity and efficiency.

Among the lowest loss architectures currently available is the combination of telecommunication band photons with silicon photonics~\cite{Singal2017}.
Not only are silicon devices at the forefront of high performance photonic integration, but they also lead the technologically important \ac{CMOS} field.
These mature silicon industries can be leveraged to realise \ac{VLSI} photonic circuits by using on-chip silicon-based components such as ultra-high quality factor cavities~\cite{Panuski2020}, low-loss waveguides~\cite{Liu2021}, phase-shifters~\cite{Edinger2021,Gyger2021}, modulators~\cite{Chakraborty2020} and near-unity single photon detectors~\cite{Gyger2021}.
For these reasons, realising telecom-wavelength emitters in silicon for \ac{QIP} applications is a highly active area of research, with artificial atoms showing remarkable progress as optically active centres~\cite{Redjem2020,Prabhu2023,Redjem2023,Hollenbach2022,Bergeron2020,Higginbottom2022,Higginbottom2023}.

Realising a high performance spin-photon interface imposes two additional requirements on optically active centres in silicon: long electron spin coherence times and long optical coherence times~\cite{Kimble2008,Duan2004,McAuslan2009}.
For this purpose, \erthree ions are strong candidates.
\erthree ions possess an electrically shielded intra-shell optical transition with emission in the lowest loss telecommunication C-band and an effective \spinhalf ground electron spin state transition.
Important \ac{QIP} milestones have been achieved using \erthree ions in insulator host crystals.
For example, \erthree ensembles in \ce{Y_2SiO_5} exhibit optical linewidths down to \SI{73}{\hertz} in large magnetic fields~\cite{Bottger2009}, and nuclear spin coherence times of over one second~\cite{Rancic2018}.
Furthermore, ground state electron spin coherence times of up to \SI{23}{\milli\second} have been measured in \erthree:\ce{CaWO_4}~\cite{Dantec2021}.
\erthree ensembles have also been used to demonstrate prototype microwave-to-optical transducers~\cite{Williamson2014,Fernandez2015,Fernandez2019}, including integrated on-chip geometries~\cite{Rochman2023}.

At a single ion level, \erthree ions in \CaWO have demonstrated indistinguishable single photon emission at telecommunication wavelengths~\cite{Ourari2023}, enabled by \SI{150}{\kilo\hertz} optical linewidths and \num{12}-hour long spectral diffusion of \SI{63}{\kilo\hertz}.
In addition, this study achieved \SI{>97}{\percent} single-shot readout fidelity optical quantum nondemolition measurements, similar to observations in \erthree:\ce{Y_2SiO_5}~\cite{Raha2020}.
By embedding \erthree ions in \ce{LiNbO_3}, the emission of single ions can also be controlled~\cite{Yang2023}.
Finally, the large gyromagnetic moment of the \erthree electron spin has enabled single \ac{ESR} detection~\cite{Wang2023}.

Despite the individual appeal of both silicon as a host material platform and \erthree ions as a spin-photon interface, \erthree:\ce{Si} systems have not yet demonstrated comparable performance to \erthree ions in insulators.
While the optical properties have been studied, resulting in the observation of sub-megahertz homogeneous linewidths~\cite{Berkman2023,Gritsch2022}, there are few studies on the \erthree electron spin transition in silicon or other semiconductor hosts. 
To date, the \erthree:\ce{Si} electron spin coherence time has solely been measured using bulk \ac{ESR}, demonstrating a \ttwohahn of \SI{7.5}{\micro\second} in natural silicon, which is likely limited by superhyperfine coupling to \sitwonine spins~\cite{Hughes2021}.
Realising long \erthree:\ce{Si} electron spin coherence times on optically active sites has therefore remained elusive.

In this paper, we use optical detection to investigate both the electron spin and optical transition properties of two \erthree sites within a \puresi (\SI{<0.01}{\percent} \sitwonine) crystal.
We observe, for the first time in any semiconductor, long \erthree electron spin coherence times of \SI{0.8}{\milli\second} and \SI{1.2}{\milli\second}.
In addition, we measure inhomogeneous linewidths of \SI{148}{\mega\hertz} and \SI{72}{\mega\hertz} as well as homogeneous linewidths below \SI{66}{\kilo\hertz} and \SI{52}{\kilo\hertz}, respectively.
Our combined observed optical and spin measurements and the excellent \ac{VLSI} potential of the silicon host present \erinsi as a promising candidate for a broad range of \ac{QIP} applications.

\section*{Photoluminescent excitation spectroscopy}

We first characterise \erthree sites in a \puresi sample using \ac{PLE} spectroscopy in a similar approach to Ref. \cite{Berkman2023} using the experimental setup shown in Fig. \ref{fig:PLE}(a) with no applied \ac{RF} field. 
The \erinsi sample is sandwiched between a fibre ferrule and a \ac{SSPD}.
We package two of these sandwiched devices, each containing an \erinsi sample from the same wafer, and cool down one device to \SI{20}{\milli\kelvin} in a dilution refrigerator and the other to \SI{300}{\milli\kelvin} in a \hethree system.
After calibrating the laser power to ensure consistent optical excitation power across the sampled frequency range, we pulse the laser at a fixed frequency and monitor the laser wavelength using a Bristol \num{621}A wavemeter.
The resulting \ac{PL} counts on the \ac{SSPD} are recorded from \SI{20}{\micro\second} up to \SI{200}{\micro\second} after each pulse (Methods). 
Resonances are observed from \SIrange{193.4}{197.0}{\giga\hertz} for the sample in the dilution refrigerator, as shown in Fig. \ref{fig:PLE}(b).
The large number of resonances arises from the \erthree ions occupying a combination of substitutional or interstitial \erinsi sites~\cite{Kenyon2005, Przybylinska1996}, with multiple lines resulting from excitation to different crystal field levels (Fig. \ref{fig:PLE}(c)).

To isolate lines possessing a long electron spin \tone, a prerequisite for a long coherence time, a \SI{10}{\milli\tesla} magnetic field is applied along the [110] crystallographic axis (Fig. \ref{fig:PLE}(a)) to lift the spin degeneracy in each crystal field level (Fig. \ref{fig:PLE}(c)). 
For all of the sites we observe here, the ground states form doublets and can be treated as an effective spin-half system with states \spindown and \spinup. 
If \tone of the spin states is sufficiently long, optically induced spin flips hyperpolarise \erthree ions to non-resonant ground states, e.g., into the \spinup state when exciting transition \num{1} in Fig. \ref{fig:PLE}(c). This reduces the \ac{PL} signal compared to that observed at zero magnetic field.

The \ac{PL} intensity can be recovered if a second laser excites the \erthree ions that are shelved into the hyperpolarised state, e.g., transition 2 in Fig. \ref{fig:PLE}(c) for a hyperpolarised \spinup state. 
The excitation frequencies under an applied magnetic field (Fig. \ref{fig:PLE}(c)) can be determined by plotting the \ac{PL} signal as function of the two laser frequencies, which produces a characteristic pattern shown in Fig. \ref{fig:PLE}(d) for the \SI{1532.853}{\nano\metre} site (\Sitec) at $B = \SI{10}{\milli\tesla}$.
Here, bright Zeeman resonances appear at laser frequencies of $\pm(f_1-f_0)=\SI{\pm175}{\mega\hertz}$ and $\pm(f_0+f_1)=\SI{\pm640}{\mega\hertz}$ detuned from the centre of the resonance at \SI{0}{\milli\tesla}, giving $\Delta f_0/\Delta B=\SI[per-mode=repeated-symbol]{46.5}{\giga\hertz\per\tesla}$ and $\Delta f_1/\Delta B=\SI[per-mode=repeated-symbol]{81.5}{\giga\hertz\per\tesla}$. 
Sites with symmetry below the T$_d$ symmetry of the crystal will have multiple orientations that may be non-degenerate in the applied field.
For \Sitec, a second orientation with $f_0 = f_1 = \SI{350}{\mega\hertz}$ is visible: this orientation gives rise to the bright horizontal and vertical \ac{PL} lines at zero frequency detuning and the signals near the 3\&4 and 4\&3 peaks of the first orientation (dotted black circles in Fig. \ref{fig:PLE}(d)).
The different crystal field levels belonging to a single site (Fig. \ref{fig:PLE}(b)) can be identified by a modification of the above method (Methods).

\section*{Optical transition measurements}
We choose two of the brightest identified sites that exhibit well-distinguished lines under a magnetic field (\Sitec and \Sited) for further investigation (Methods).
The inhomogeneous lines of these sites are displayed in Fig. \ref{fig:optical}(a) and fit a Lorentzian lineshape, which is consistent with previous studies~\cite{Weiss2021,Gritsch2022,Berkman2023}. 
\Sitec shows consistent \ac{FWHM} of the inhomogeneous lines of around \SI{ 150}{\mega\hertz}, whereas the linewidths of the different crystal field transitions of \Sited vary, with \ac{FWHM} values of \SI{145}{\mega\hertz}, \SI{72}{\mega\hertz} and \SI{575}{\mega\hertz}.
We attribute the larger linewidth of the third crystal-field level excitation, compared to the excitations to the lower crystal-field level, to homogeneous broadening arising from a fast decay from the third, to the first and second crystal-field level in the \thirteenhalf manifold. 
The inhomogeneous linewidths seen here, below \SI{150}{\mega\hertz}, are comparable to the narrowest linewidths observed in bulk \erthree in insulating systems~\cite{Macfarlane1992} and are less than half the width of any inhomogeneous lines observed to date for thin-region implanted \erinsi~\cite{Berkman2023,Gritsch2022}.

We use spectral holeburning on the transition to the lowest \thirteenhalf crystal field level to investigate the homogeneous linewidths of the two sites.
Our method is based on the technique presented in Ref. \cite{Berkman2023}, which relies on the saturating behaviour of an atomic transition as function of laser power, i.e., $I(2P_{sb})<2I(P_{sb})$, where $I$ is the \ac{PL} intensity and $P_{sb}$ is the optical power on the sample of one laser frequency sideband.
We make two modifications to this technique to enhance the \ac{SNR}: the pump and probe pulse occur simultaneously, and this pump-probe doublet is repeated every $\Delta f_\text{comb}= \SI{3.4}{\mega\hertz}$ over the inhomogeneous peak, resulting in a frequency comb (Fig. \ref{fig:optical}(b)). 
When the pump-probe separation $f_\text{doublet}$ is zero, the total \ac{PL} intensity is $N\cdot I(2P_{sb})$ (dashed red line in Fig. \ref{fig:optical}(b)) where $N$ is the total number of ions excited, whereas in the case when $f_\text{doublet}$ exceeds the homogeneous linewidth ($\gamma_h$), the total \ac{PL} intensity is given by $2N\cdot I(P_{sb})$ (solid red line in Fig. \ref{fig:optical}(b)).
Because the spectral comb generated is similar for $\Delta f_\text{doublet}$ and $-\Delta f_\text{doublet}$, the spectral hole can be mapped by sweeping $\Delta f_\text{doublet}>0$ and mirroring the data along $\Delta f_\text{doublet}=0$.

Fig. \ref{fig:optical}(c) shows the resulting spectral holes for each site for a \SI{50}{\micro\second} pulse at the lowest optical power set by the measurement \ac{SNR} ratio, i.e., $P_{sb} = \SI{139}{\nano\watt}$ and $P_{sb} = \SI{35}{\nano\watt}$ for \Sitec and \Sited, respectively. 
Because the measured spectral hole is the convolution of two burnt spectral holes (Fig. \ref{fig:optical}(b)), $\gamma_h$ is equal or less than \num{2} times the hole \ac{FWHM}~\cite{moerner1988}. 
Lorentzian hole fits gives homogeneous linewidth upper bounds of \SI{66}{\kilo\hertz} and \SI{52}{\kilo\hertz} for \Sitec and \Sited, respectively.
The observed spectral hole widths are close to the system spectral resolution limit of approximately \SI{30}{\kilo\hertz} set by the laser intrinsic linewidth (\SI{10}{\kilo\hertz}) and the Fourier limited broadening of the used optical pulses (\SI{20}{\kilo\hertz}).
Future experiments could hence utilise a more stable laser source to obtain narrower bounds on $\gamma_h$.
Nonetheless, the optical properties of these sites are among the narrowest \erinsi linewidths observed thus far~\cite{Gritsch2022,Berkman2023,Weiss2021}.
 
\section*{Spin transition measurements}
The spin transition was studied via \ac{ODMR} by hyperpolarizing the spin states into the \spindown state, applying an \ac{RF} pulse to partially repopulate the \spinup state, and using state-selective optical excitation of the \spinup state (transition 2 in Fig. \ref{fig:PLE}(c)) to measure the \erthree population in the \spinup state. 
The static magnetic fields of \SI{12.37}{\milli\tesla} for \Sitec and \SI{10.56}{\milli\tesla} for \Sited, are chosen to match the Zeeman splitting of the site (\SI[per-mode=repeated-symbol]{46.5}{\giga\hertz\per\tesla} for \Sitec, and \SI[per-mode=repeated-symbol]{64.7}{\giga\hertz\per\tesla} for \Sited) to the \ac{RF} antenna resonance frequency: \SI{575}{\mega\hertz} for \Sitec (\hethree system), and \SI{683}{\mega\hertz} for \Sited (dilution refrigerator). 

Decreasing the \ac{RF} power results in narrower spin inhomogeneous linewidths, indicating power broadening at higher \ac{RF} powers. 
For \Sitec, a \ac{FWHM} of \SI{282}{\kilo\hertz} is observed at \SI{-10}{\decibelm} \ac{RF} power to the antenna (Fig. \ref{fig:ODMR}(a)). 
This site shows a long term drift of \SI{\pm 0.5}{\mega\hertz} over minutes.
The drift to higher frequencies likely indicates the sample holder moving in the field of the (persistent-mode) superconducting magnet in the \hethree system.
The corresponding linewidth of \Sited at an equivalent \SI{-10}{\decibelm} of \ac{RF} power is \SI{129}{\kilo\hertz} (Fig. \ref{fig:ODMR}(a)), with no observable long term drift, attributed to a more stable setup in the dilution refrigerator, allowing the observation of a narrower linewidth.

Rabi oscillations are obtained by varying the \ac{RF} pulse length with the frequency tuned to the centre of the inhomogeneous spin line. 
Here, we focus on \Sited due to the better long term stability of that setup.
The Rabi oscillations are fit with
\begin{equation}
    P_{0\rightarrow 1} = A[1-\exp{(-t/T_{2,\text{Rabi}})}]\cos{(2\pi f_\text{Rabi} t)},
\end{equation}
\noindent where $A$ is the amplitude, $t$ is the pulse length, $T_{2,\text{Rabi}}$ is the Rabi decay constant and $f_\text{Rabi}$ is the Rabi frequency. 
For \SI{-16}{\decibelm} \ac{RF} power to the antenna, $f_\text{Rabi}=\SI{70}{\kilo\hertz}$ and $T_{2,\text{Rabi}}=\SI{50}{\micro\second}$, while at \SI{12}{\decibelm} power, $f_\text{Rabi}$ increases to \SI{1.8}{\mega\hertz} and \ttworabi decreases to \SI{2.2}{\micro\second} (Fig. \ref{fig:ODMR}(b)). 
Pulse widths shorter than \SI{350}{\nano\second} were not applied due to limitations on the \ac{RF} source, as well as the \SI{15}{\mega\hertz} bandwidth of the \ac{RF} antenna.

The high Rabi frequency seen here is comparable with other state-of-the-art electron spin qubit systems in silicon~\cite{Pla2012,Fricke2021,Philips2022,Bergeron2020}, and indicates fast spin-flip operations are possible for \erthree ions in silicon. 
The Rabi frequencies scale linearly with the applied electric field, i.e. $f_\text{Rabi}\propto\sqrt{P}$, rather than showing a saturation of the Rabi frequencies at these powers, which has been observed in semiconductor spin qubits at high \ac{RF} powers~\cite{Vahapoglu2022,Undseth2023,Yoneda2018,Yoneda2014,Takeda2016,Nakajima2020}.
Hence, higher Rabi frequencies are expected to be observable by optimising the antenna geometry to allow higher magnetic fields with less heat dissipation.

A Hahn echo sequence~\cite{Hahn1950} is used to measure the decoherence time (\ttwohahn) of \Sitec and \Sited. 
To optically detect the resulting refocused \erthree spins, we apply an additional \pitwox (or -\pitwox) pulse after the refocusing to project the spin states in the \spinup (\spindown) state and then use state-selective excitation to probe the spin state population (Fig. \ref{fig:PLE}(c)). 
We record the signal for varying time $\tau$ for two sequences (final \pitwox or final -\pitwox) and subtract the two signals to remove background counts, giving a signal decay depending only on \ttwohahn (Fig. \ref{fig:ODMR}(d)). 
This trace is fit with:
\begin{equation}
    I = A\exp{[-(\tau/T_{2,\text{Hahn}})^x]},
\end{equation}
where $A$ is the fluorescence amplitude, $T_{2,\text{Hahn}}$ is the exponential decay constant and $x$ is the stretch factor that can indicate the dominant decoherence mechanism~\cite{Hughes2021}.
The decay curves for \Sitec and \Sited give a \ttwohahn of \SI{0.78}{\milli\second} and \SI{1.19}{\milli\second}, and stretch factors of \num{3.80} and \num{3.34}, respectively. 
This remarkable \num{100}-fold improvement compared to the previously reported \SI{7.5}{\micro\second} coherence time is, as suggested in Ref. \cite{Hughes2021}, attributed to superhyperfine coupling to the \sitwonine spins, which we have confirmed by measuring the coherence time of \Sited in a natural silicon sample (Supplementary Information).
Similar stretch factors have also been observed in other centres in silicon, including phosphorus\cite{Chiba1972,Tyryshkin2006} and T centres~\cite{Bergeron2020}, which have been attributed to a phase mismatch between the \ac{RF} field and the electron spin precession frequencies, consequently leading to a shorter observed \ttwohahn.
In future \erinsi studies, a potential higher bound on the \ttwohahn can be extracted by mitigating instrumental phase noise, or when conducting \ac{ESR} measurements, by taking the amplitude of the \ac{ESR} signal~\cite{Tyryshkin2006}.

\section*{Conclusion and outlook}

In this study, we investigated \erthree ions in \puresi at two sites and observed narrow optical linewidths and millisecond ground electron spin coherence times in both sites.
These long coherence times indicated that \erinsi meets the fundamental requirements of a versatile telecom-compatible spin-photon interface in silicon.
Furthermore, because the measurements were carried out on near-surface \erthree ions implanted at a depth from \SIrange{200}{700}{\nano\metre}, the findings of this study approaches the \erinsi environment in more complex devices which utilise near-surface ions.
Here, the \ac{VLSI} potential of silicon can be harnessed to fabricate photonic devices such as \erinsi implanted \acl{SOI} nanophotonic cavities in order to enhance the emission rate or achieve non-linear effects~\cite{Gritsch2023,Duan2004,McAuslan2009}. 
Future \erinsi research can also benefit from the increasingly accessibility of \puresi or \puresi-on-insulator wafers, which are realised through new approaches such as high fluence \puresi ion implantation~\cite{Holmes2021} or \acl{MBE} growth~\cite{Liu2022}.

We showed that the optical homogeneous linewidth of both sites is less than \SI{70}{\kilo\hertz}. 
This itself is already narrow, confirming these sites as good optical quantum information candidates, but the prospect of a much narrower true linewidth motivates additional study to extract the true linewidth.
Photon echo methods, while challenging for such small ensembles, can provide the required accuracy at the same time as allowing another important parameter, the spectral diffusion, to be measured.
The current measurements can only limit the spectral diffusion on timescales less than the pulse length (\SI{50}{\micro\second}) to be less than \SI{70}{\kilo\hertz}.
\Ac{QIP} applications for \erinsi, such as sources of indistinguishable photons, typically require medium or long-term stability in the optical frequency and, therefore, low spectral diffusion.
This is a challenge for many optical centres, particularly in nanophotonic structures where surface states and other fabrication effects lead to both line broadening and spectral diffusion on short and long timescales~\cite{Sangtawesin2019,Macquarrie2021}.
Fortunately, \erthree lines tend to remain narrow and stable because their low electric field susceptibility generates a weak sensitivity to noise and defects~\cite{Ourari2023,Gritsch2022}.

These \erthree sites can be compared to the other optical spin photon candidate in a telecom band, the silicon T centre, a complex carbon-related radiation defect with a \SI{1326}{\nano\metre} transition in the O-band~\cite{Bergeron2020,Higginbottom2022}.
In \puresi, both sites have comparable electron spin coherence times of \SI{2.1}{\milli\second} (T centre) and \SI{1.2}{\milli\second} (\erthree), and optical inhomogeneous linewidths of \SI{33}{\mega\hertz}~\cite{Bergeron2020} (T centre) and \SI{72}{\mega\hertz} (\erthree). 
Their other optical properties place them in different, complementary regimes: the T centre has a stronger oscillator strength of $10^{-4}$ and an associated \si{\micro\second} short lifetime~\cite{Bergeron2020,Dhaliah2022}, and broad homogeneous line of \SI{690}{\kilo\hertz}~\cite{DeAbreau2022} with \SI{27}{\mega\hertz} of spectral diffusion~\cite{Macquarrie2021}, while \erthree has a weaker oscillator strength (not measured in silicon, but $10^{-7}$ is typical~\cite{Liu2005}), \SIrange{0.1}{1.6}{\milli\second} lifetimes~\cite{Gritsch2022,Berkman2023} and very narrow homogeneous lines below \SI{70}{\kilo\hertz}. 
Both optical centres are in similar, initial stages of study, but already show significant appeal for spin-photon interfaces. 
Further development is likely to improve key properties and elucidate the regimes and applications best suited to each centre.

This work shows for the first time that \erinsi meets the spin and optical coherence time requirements for a spin-photon interface.
Taken together with the unmatched mature fabrication technology existing in silicon, these properties present \erinsi as a promising material of high interest for realising a scalable, telecommunication-compatible platform for quantum computing and quantum network applications.

\section*{Methods}

\subsection*{Implant and anneal recipe}

The two \puresi samples both originate from the material used for the study in Ref. \cite{Kobayashi2021}, which contains a background doping of \ce{B} at approximately \SI{1e15}{\centi\metre\tothe{-3}}.
To study the optical transitions without the complication of hyperfine structure, the nuclear-spin-free \ce{^{170}Er} isotope is implanted with multiple ion energies and fluences into one side of the \SI{490}{\micro\metre} double-side-polished \puresi wafer to form a uniform concentration profile of \SI{1e16}{\centi\metre\tothe{-3}} over a depth of \SI{200}{\nano\metre} to \SI{700}{\nano\metre}. 
Following implantation, the wafers are consecutively diced and annealed at \SI{700}{\celsius} for \SI{10}{\minute} in an \ce{N_2} atmosphere, which is known to lead to optically active \ce{Er^{3+}} ions in silicon~\cite{Yin2013, Weiss2021, Gritsch2022, Berkman2023}. 

\subsection*{Photoluminescent excitation spectroscopy}

The emission of the \erthree ions is collected by an in-situ \ac{SSPD} as presented in Ref. \cite{Berkman2023}.
Here, a ferrule containing a CCC1310-J9 fibre was polished and used to sandwich the sample between the \ac{SSPD} and the ferrule. 
To obtain the \ac{PLE} spectrum, the laser intensity is swept from \SIrange{190.70}{197.36}{\tera\hertz} (\SIrange{1572}{1519}{\nano\metre}) in steps of \SI{20}{\mega\hertz}.

In addition to this in-situ detection configuration, we add a single-layer coiled copper wire antenna between the sample and the fibre ferrule for the \ac{RF} control of the electron spins.
The inner diameter of the copper coil is larger than the mode diameter of the laser light emitted from the fibre end, thus providing an unobstructed path for the laser light to reach the sample.
This copper wire is present, but unused during the \ac{PLE} scan.

One \puresi sample is situated in a dilution refrigerator at a temperature of \SI{20}{\milli\kelvin}, and one \puresi sample in a \hethree system at a temperature of \SI{300}{\milli\kelvin}.
These low temperatures are to ensure a low dark count rate on the \ac{SSPD}, and long spin-lattice and spin-spin relaxation times. 
Moreover, these temperatures are low enough to minimise non-radiative recombination of the \ce{Er^{3+}} optical transition~\cite{Palm1996,Taguchi1998,Priolo1998}.

To excite the \ce{Er^{3+}} ions, we use a semiconductor diode laser (Pure Photonics PPCL550) with an output pulse modulated by two \acp{AOM} connected in series, resulting in an extinction ratio greater than \SI{100}{\decibel}, similar to the excitation schematic in Ref. \cite{Berkman2023}.
After the excitation pulse, we record the number of counts from the \ac{SSPD} for \SI{200}{\micro\second} with a digital counter (Keysight 53131A or National Instruments PCI-6602).
The \ac{SSPD} current bias is set to zero during the excitation pulse and is reset to its nominal value \SI{20}{\micro\second} after the excitation pulse.
This \ac{SSPD} blanking mitigates the current-induced Joule heating, hence allowing a faster reset time after the excitation pulse is extinguished~\cite{Berkman2023}.
For this reason, the counts in the first \SI{20}{\micro\second} after the excitation pulse are ignored in all \ac{PLE} measurements.

\subsection*{PLE spectrum site extraction}

The resonances in the \ac{PLE} spectrum belonging to the same site can be identified with a modification of the two-colour excitation sequence used to recover the Zeeman spectrum: instead of shifting the frequency of the second laser by a small amount to excite one of the transitions to the same excited state Zeeman doublet, it is shifted to the centre of a different line in the \ac{PLE} spectrum. 
If this causes the recovery of the signal, that line is a higher energy crystal field line belonging to the same site (Extended Data Fig. 1 Site F crystal field measurement), and by testing all lines in the spectrum, all visible transitions associated with one site can be determined.
This procedure is performed on the \num{7} brightest resonances exhibiting long-lived spin states, identifying them as arising from seven sites with long electron spin lifetimes (Fig. \ref{fig:PLE}(b) and Extended Data Table 1 Resonances with long-lived electron spin states).
Of these sites, \Sited exhibits the brightest resonance within the \ac{PLE} spectrum at \SI{195571}{\giga\hertz}, and is therefore the first resonance we used for spin coherence time measurements. 
We furthermore choose \Sitec as a second resonance to measure to confirm that the characteristics of \Sited were not solely limited to one site. 
\Sitec is the third-brightest and produced the clear Zeeman pattern shown in Fig. \ref{fig:PLE}(d); Site C, the second brightest, is excluded due to its unclear Zeeman pattern indicating a large number of site orientations, which results in a lower intensity per Zeeman transition.

\subsection*{Optically detected magnetic resonance spectroscopy}

To control the electron spin states, we use \ac{RF} pulses from the copper coil generated by a Keysight N5182B vector source.
In order to mitigate the heat load of the \ac{RF} antenna, we introduce an $LC$ circuit to enhance the \ac{RF} emission at a desired resonance frequency.
We set the resonance frequency at approximately \SI{600}{\mega\hertz}, which is high enough to distinguish the two \mszero optical transitions as seen in Fig. \ref{fig:PLE}(d), hence allowing state-selective optical excitation.

The $LC$ circuit is formed by adding two capacitors to the copper wire antenna: one capacitor is connected in series to set the resonance frequency of the antenna, and the other capacitor is connected in parallel to match the impedance.
The resulting resonance frequency is measured by sweeping the reflection of the antenna as function of the \ac{RF} frequency using a \SI{9}{\giga\hertz} Keysight FieldFox.
The resonance frequency of the antenna in the dilution refrigerator is \SI{683}{\mega\hertz} at a bandwidth of \SI{15}{\mega\hertz}, and the resonance frequency of the antenna in the \hethree system is \SI{590}{\mega\hertz} with a bandwidth of \SI{20}{\mega\hertz}.

\subsection*{Cyclicity measurements}

\begin{figure}[t]
    \centering
    \includegraphics[]{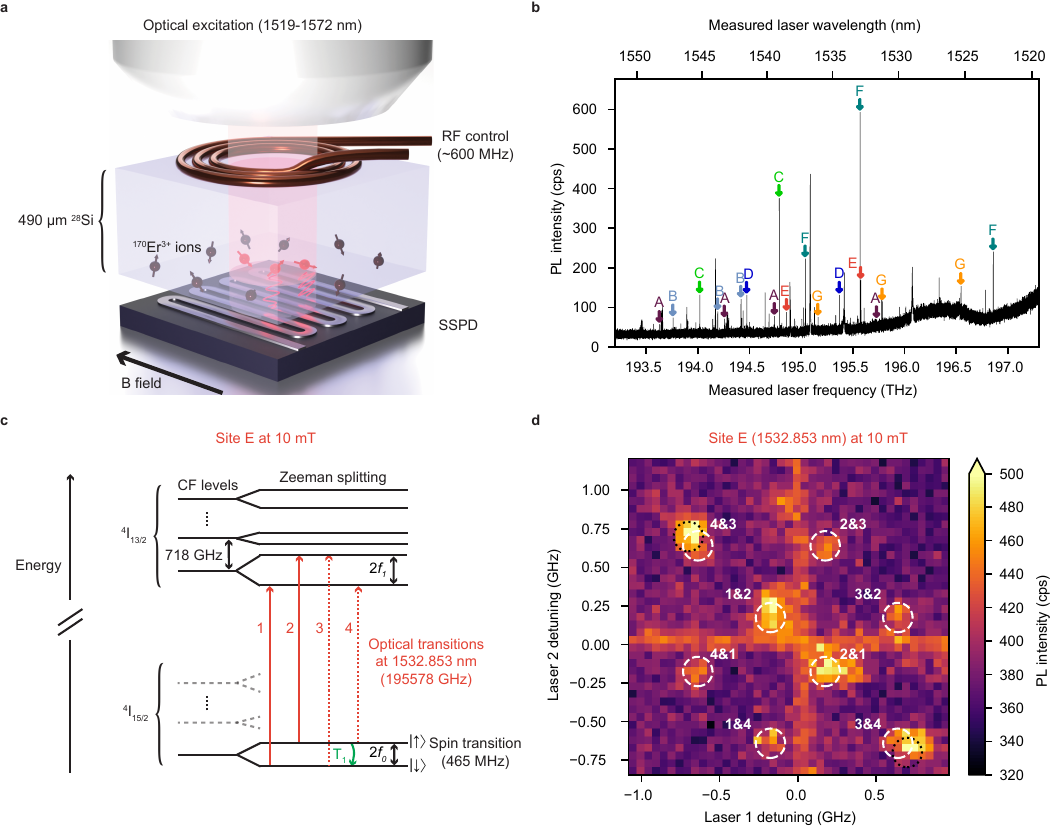}
    \caption{Photoluminescent excitation spectrum of \erthree ions in silicon. (\textbf{a}), Schematic of the setup to perform \acf{PLE} and \acf{ODMR} spectroscopy of \erthree ions in silicon. The optical excitation consists of a laser pulse which excites the \erthree ions resonant with the laser. Directly after the pulse, the \acf{PL} of these \erthree ions are collected by a \acf{SSPD}. The antenna consists of a copper wire and is solely used for the \ac{ODMR} measurements. 
    (\textbf{b}), \ac{PLE} spectrum of the \erthree ions in \puresi. 
    The letters denote the different sites observed and the resonances that belong to the same site.
    (\textbf{c}) Energy level diagram of \Sitec under \SI{10}{\milli\tesla}. The optical transition is excited using the laser light, whereas the spin transition is controlled using the \ac{RF} antenna in (a). The values in this diagram are extracted from the \ac{PLE} spectrum in (a), as well as the Zeeman spectrum in (d). The solid red lines refer to spin-conserving transitions, while the dashed red lines refer to spin-flipping transitions.
    (\textbf{d}) 2D Zeeman spectrum, showing a recovery (dashed circles) of the signal solely when the two lasers excite from both of the ground electron spin states as displayed in (c). The numbers next to each recovered signal correspond, respectively, to the first and second laser, and the transitions excited in (c). The detuned frequencies where the Zeeman signals are observed occur at $\pm(f_1-f_0)=\SI{\pm175}{\mega\hertz}$ and $\pm(f_0+f_1)=\SI{\pm640}{\mega\hertz}$. An additional \erthree orientation is visible with $f_0 = f_1 = \SI{350}{\mega\hertz}$, displayed by the bright horizontal and vertical line, and the bright signal close to transitions \num{4}\&\num{3} (\num{3}\&\num{4}), indicated with the black dashed circles.
}
    \label{fig:PLE}
\end{figure}

\begin{figure}[h]
    \centering
    \includegraphics[]{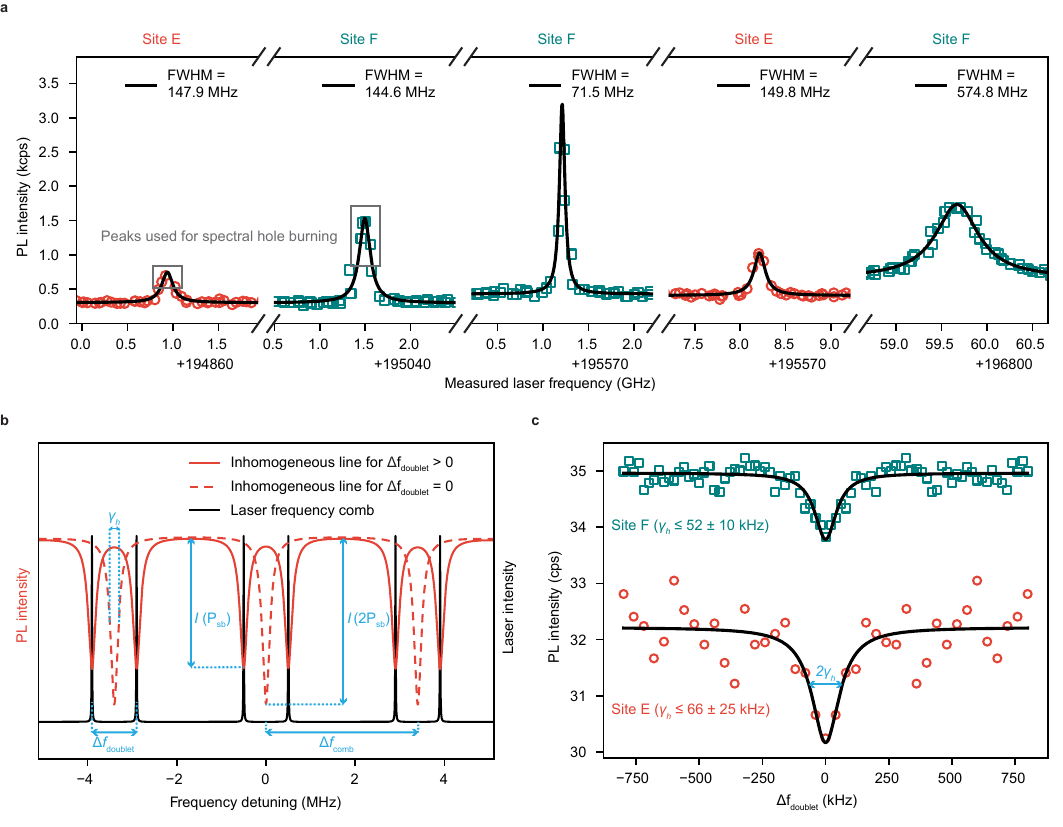}
    \caption{Optical transition properties of \Sitec and \Sited. \textbf{(a)}, \Acf{PLE} spectrum of \Sitec and \Sited, fit with a Lorentzian distribution to extract the inhomogeneous linewidths, showing narrow linewidths for the sites. \textbf{(b)}, Schematic of the spectral hole burning technique to extract the upper bound on the homogeneous linewidth ($\gamma_h$). For \Sitec and \Sited, a frequency comb of $N=\num{18}$ doublets is used, with a $\Delta f_\text{comb}$ of \SI{3.4}{\mega\hertz}. $P_{sb}$ refers to the optical power on the sample for each sideband. At $\Delta f_\text{doublet} = 0$, $N$ spectral holes are burned, whereas for $\Delta f_\text{doublet}>\gamma_h$, $2N$ spectral holes are burned. Because of atomic saturation, the \acf{PL} intensity increases for increasing values of $\Delta f_\text{doublet}$, allowing the extraction of the upper bound on $\gamma_h$. \textbf{(c)}, \ac{PL} intensity as a function of $\Delta f_\text{doublet}$ for \Sitec ($P_{sb} = \SI{139}{\nano\watt}$) and \Sited ($P_{sb} = \SI{35}{\nano\watt}$). The resulting spectral holes are fit with a Lorentzian distribution, where the $\gamma_h$ is less than half the \ac{FWHM}.}
    \label{fig:optical}
\end{figure}

\begin{figure}[h]
    \centering
    \includegraphics[]{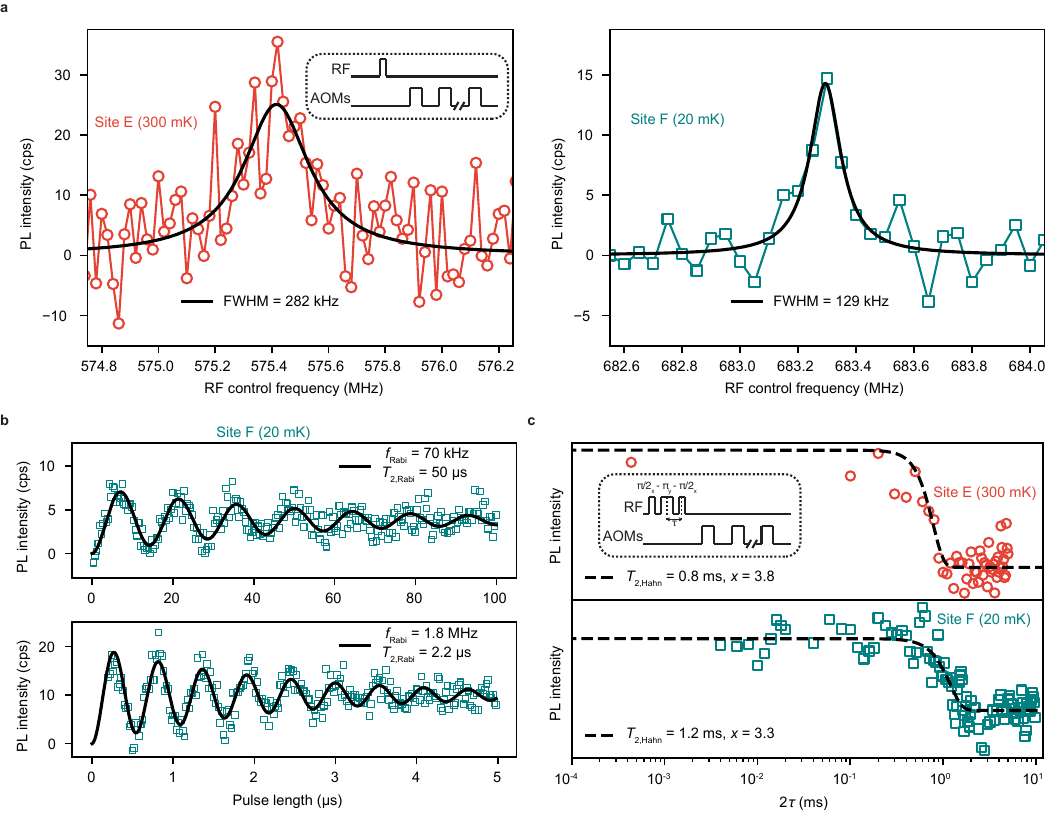}
    \caption{Spin transition properties of \Sitec and \Sited. \textbf{(a)}, \Acf{ODMR} signal of the spin transition as function of the \ac{RF} frequency of the antenna at \num{-10} dBm \ac{RF} power to the antenna. The static magnetic fields applied are \SI{12.37}{\milli\tesla} and \SI{10.56}{\milli\tesla} for \Sitec and \Sited, respectively. Both data are fit with a Lorentzian distribution. \textbf{(a,inset)}, The pulse sequence used for obtaining the \ac{ODMR} signal.
    \textbf{(b)}, \Ac{ODMR} signal of \Sited at different \acf{RF} pulse lengths, using the same sequence as in (a, inset). 
    \textbf{(c)}, Hahn echo decay traces as function of delay between the pulses ($\tau$). Both \Sitec and \Sited show long electron spin coherence time at \SI{300}{\milli\kelvin} and \SI{20}{\milli\kelvin}, respectively. \textbf{(c,inset)}, The Hahn echo pulse sequence used. Here, a final \pitwox (-\pitwox) is applied to project the spin states to the \spinup (\spindown) state. This allows optical state-selective excitation to extract the \ac{ODMR} signal.
    }
    \label{fig:ODMR}
\end{figure}

We measure the number of optical pulses required to depolarise a spin state, depending on the cyclicity of the optical transition excited.
During the \ac{ODMR} sequence, a number of optical pulses are applied after the \ac{RF} pulse for two reasons: to initialise the spin state and to enhance the \ac{SNR} of the collected \ac{PL}.
The number of optical pulses after the \ac{RF} sequence is chosen such that $N_\text{pulses} > N_\text{cyclicity}$, where $N_\text{pulses}$ is the number of pulses after the \ac{RF} sequence and $N_\text{cyclicity}$ the number of optical pulses needed to depolarise the spin states, depending on the optical pulse length and power specific to the measurements.
After every optical pulse in this state-selective optical sequence, the \ac{PL} is collected.
For our measurement, we excite the \mszero transitions (red solid lines in Fig. \ref{fig:PLE}(c)).

Here, we show the cyclicity measurement on \Sitec, however, the same procedure was applied on \Sited which contains a relatively equal cyclicity.
For this measurement, an \ac{EOM} was used to rapidly change the excitation frequency.
A frequency of \SI{3}{\giga\hertz} was applied on the \ac{EOM}, leading to a single carrier and two sidebands.
The carrier was suppressed by changing the polarisation of the light and by optimising the DC voltage to the \ac{EOM}.
One of the remaining two sidebands was centred on the resonance belonging to the \spindown transition for the pump pulse and the other sideband was located off-resonantly \SI{3}{\giga\hertz} at a lower frequency.
The \spindown transition is depolarised by exciting this transition for \SI{1}{\milli\second} at \SI{120}{\micro\watt} (\SI{60}{\micro\watt} per sideband) of optical power on the sample.
Afterwards, the frequency of the sideband is set to the transition in resonance with the \spinup transition to initiate the probe sequence.
The probe sequence consists of \num{100} excitation pulses of \SI{10}{\micro\second} at the same optical power on the sample, which are spaced \SI{200}{\micro\second} apart.
The \ac{PL} is collected in the \SI{200}{\micro\second} bins, of which we ignore the first \SI{20}{\micro\second} due to the \ac{SSPD} recovery time as explained in Methods.
This sequence is repeated \num{200} times and the \ac{PL} at each pulse number is averaged (red curve in Fig. \ref{fig:Cyclicity_E}).
This sequence was repeated with the pump pulse off-resonantly centred \SI{200}{\mega\hertz} from the \spindown transition (black curve in Fig. \ref{fig:Cyclicity_E}).
As expected, the count rate of the on-resonant curve is higher than the off-resonant curve due to \ac{PL} excitation of the hyperpolarised \spindown state, and eventually reaches the same \ac{PL} intensity after the \spindown state is depolarised.
Fitting an exponential to the difference of these two curves results in a $1/e$ value of \num{30} pulses.
During the \ac{ODMR} measurements, the same optical parameters of \SI{10}{\micro\second} pulse length and \SI{60}{\micro\watt} of optical power were used to depolarise and read out the spin states.

\begin{acknowledgments}  
The \ac{SSPD} fabrication is performed at the NSW Node and ACT Node of the NCRIS-enabled Australian National Fabrication Facility.
We acknowledge the AFAiiR node of the NCRIS Heavy Ion Capability for access to ion-implantation facilities. 
This work is supported by the ARC Centre of Excellence for Quantum Computation and Communication Technology (Grant CE170100012) and the Discovery Project (Grant DP210101784). 
We thank Prof. M. L. W. Thewalt for providing us the \puresi wafer and we thank Dr. Sae Woo Nam for kind support to establish the \ac{SSPD} fabrication process at the University of New South Wales, Sydney. 
\end{acknowledgments}  

\clearpage

\onecolumngrid

\twocolumngrid


\FloatBarrier
\onecolumngrid
\section*{Extended data figures and tables}
\twocolumngrid
\subsection*{Extended Data Fig. 1 Site F crystal field measurement}
\setcounter{figure}{0} 
\begin{figure}[h]
    \includegraphics[trim={0 0 0 15},clip]{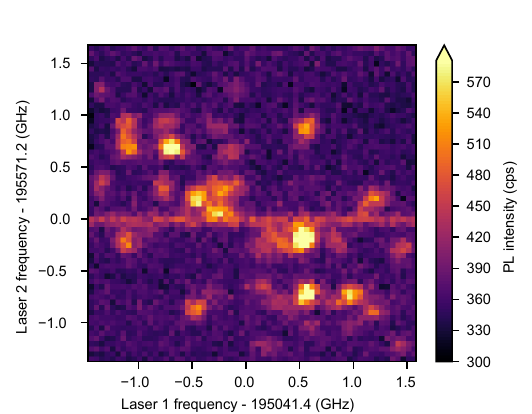}
    \caption{Zeeman measurement on \Sited at \SI{10}{\milli\tesla}, with the two lasers exciting to two different crystal field levels. The Zeeman pattern indicates that the two resonances belong to the same site, as explained in the main text. This site contains a larger number of \erthree subsite orientations compared with \Sitec (Fig. \ref{fig:PLE}(d)).}
    \label{fig:Cyclicity_E}
\end{figure}

\newpage
\subsection*{Extended Data Fig. 2 Cyclicity measurement}

\begin{figure}[h]
    \includegraphics[trim={0 0 0 0},clip]{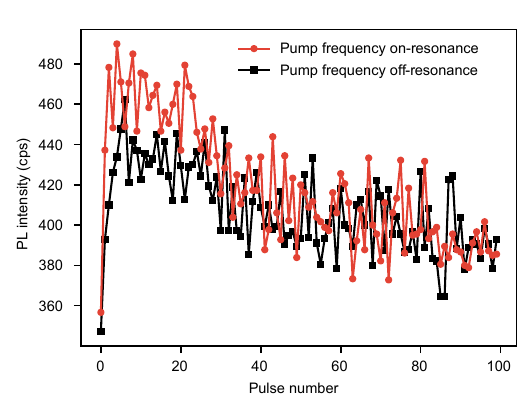}
    \caption{Cyclicity measurement on \Sitec. A pump pulse of \SI{1}{\milli\second} at \SI{60}{\micro\watt} is applied either on-resonantly on the \spindown transition, or off-resonantly \SI{200}{\mega\hertz} higher. Afterwards, \num{100} probe pulses of \SI{10}{\micro\second} at \SI{60}{\micro\watt} are applied on the \spinup transition, where the photoluminescence is collected after every probe pulse. At these optical parameters, the \ac{PL} intensity of the difference of the on-resonant and off-resonant counts drops to $1/e$ after \num{30} pulses.}
    \label{fig:Cyclicity_E}
\end{figure}

\newpage

\subsection*{Extended Data Table 1 Resonances with long-lived electron spin states}

\LTcapwidth=\columnwidth
\begin{longtable}{
>{\RaggedRight}p{0.24\columnwidth} >{\RaggedRight}p{0.1\columnwidth} 
}
    \caption{The resonances belonging to the sites in Fig. \ref{fig:PLE}(b) that show characteristics of long-lived ground electron spin states. $f_\text{centre}$ corresponds to the measured centre frequency of the resonance and \textit{Site} refers to the sites presented in Fig. \ref{fig:PLE}(b).}
    \label{tab:resonances}\\
    \thead{$f_\text{centre}$ (\si{\giga\hertz})} & \thead{Site}\\
    \midrule
        \num{193630.8} & A\\
        \num{193761.1} & B\\
        \num{194020.9} & C\\
        \num{194193.7} & B\\
        \num{194259.8} & A\\
        \num{194418.3} & B\\
        \num{194474.6} &D\\
        \num{194743.4} & A\\
        \num{194788.4} & C\\
        \num{194860.9} &E\\
        \num{195041.4} &F\\
        \num{195164.2} &G\\
        \num{195371.7} &D\\
        \num{195571.2} &F\\
        \num{195578.1} &E\\
        \num{195731.9} & A\\
        \num{195785.7} &G\\
        \num{196547.7} &G\\
        \num{196859.7} &F\\
    \bottomrule
\end{longtable}

\clearpage
\newpage
\onecolumngrid
\section*{Supplementary Information}
\twocolumngrid
\subsection*{Superhyperfine coupling of \erthree ions to \sitwonine}

To investigate the effect of \sitwonine on the decoherence of the electron spin, the Hahn echo sequence is repeated on \Sited in a natural silicon sample prepared with identical methods to the \puresi sample.
The background doping of this sample is \ce{B} at a concentration of approximately \SI{1e15}{\centi\metre\tothe{-3}}.
After implantation, the sample is annealed at \SI{700}{\celsius} for \SI{10}{\min} in a \ce{N_2} environment.
In this sample, \Sited is visible at the same frequencies of \Sited in the \puresi sample in the main manuscript.

\begin{figure}[h]
    \includegraphics[]{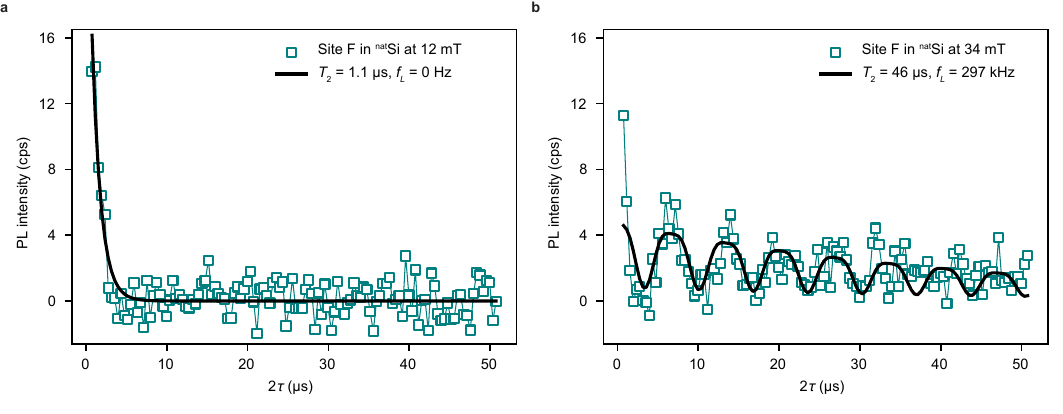}
    \caption{\textbf{(a)}, Hahn echo signal for \Sited in natural silicon at \SI{12}{\milli\tesla}. Compared with the Hahn echo signal in \puresi, the coherence time is three orders of magnitude shorter at this magnetic field due to the higher \sitwonine density. \textbf{(b)}, Hahn echo signal for \Sited in natural silicon at \SI{34}{\milli\tesla}. Here, the Hahn echo decay recovers and displays a modulated signal. The data is fit with Eq. \ref{eq:superhyperfine}, resulting in a modulation frequency equal to the Larmor frequency ($f_L$). A fast decay at shorter delays and multiple frequencies at longer delays can be observed, indicating that future site studies and measurements are required for a more accurate fit.
    }
    \label{fig:Hahn}
\end{figure}

Under \SI{12}{\milli\tesla} and a $\pi$-pulse length of \SI{500}{\nano\second}, the coherence time is \SI{1.1}{\micro\second} [Fig. \ref{fig:Hahn}(a)], which is three orders of magnitude lower than in the \puresi sample.
Considering that the \sitwonine concentration is the only significant difference compared to the \puresi sample, this shortened \ttwohahn indicates that the \sitwonine nuclear spins constitute the largest decoherence channel within natural silicon. 
This result is validated by repeating the Hahn echo measurement in a higher magnetic field, sufficient to exceed the \erthree dipolar field at distant \sitwonine spins and thus decouple these spins, leading to a recovery of the Hahn echo signal~\cite{Car2020}. 
To drive the spins at this field, we set the \ac{RF} frequency to a higher harmonic of the \ac{RF} antenna, and increase the \ac{RF} power to reach an equivalent \SI{500}{\nano\second} $\pi$-pulse length. 
This results in echo recovery at \SI{34}{\milli\tesla} showing a modulation arising from residual superhyperfine coupling.
This data is fit by assuming that the \erthree ions couple weakly to many \sitwonine nuclei \cite{Hughes2021}, giving:
\begin{equation}
    \label{eq:superhyperfine}
    I_{\tau} = I_0\exp^{-\left(\frac{2\tau}{T_2}\right)^x}\left[1-\frac{k}{4}\left[3-4\cos(2\pi f_L\tau)+\cos(4\pi f_L\tau)\right]\right], 
\end{equation}
where $I_0$ is the signal amplitude, $x$ is the stretch factor set to \num{1} since the decoherence of this site is limited by the \sitwonine nuclear spin \cite{Hughes2021,Probst2020}, $k$ is the modulation index and $f_L$ is the nuclear spin Larmor frequency at \SI{34}{\milli\tesla} (\SI{288}{\kilo\hertz}) for \sitwonine~\cite{Hughes2021,Probst2020}. 
From the fit, we obtain $k = 0.41$ and $f_L = \SI{297.3}{\kilo\hertz}$.
These values are equal to the values observed in \erthree:\ce{Si} using \ac{ESR}~\cite{Hughes2021}, and further validate that the microsecond \ttwohahn is limited by \sitwonine in natural silicon.

This fit deviates at $2\tau < \SI{2}{\micro\second}$, where a fast decay is observed, and $2\tau>\SI{30}{\micro\second}$, where multiple frequencies can be distinguished, which we ascribe to multiple \erthree subsets with different \sitwonine configurations in the environment. 
A more elaborate model could better explain these discrepancies but the finer details of the superhyperfine coupling are not within the scope of this paper.

\onecolumngrid

\clearpage
\twocolumngrid
\bibliographystyle{apsrev4-2}
\bibliography{bibliography}

\begin{thebibliography}{91}%
\makeatletter
\providecommand \@ifxundefined [1]{%
 \@ifx{#1\undefined}
}%
\providecommand \@ifnum [1]{%
 \ifnum #1\expandafter \@firstoftwo
 \else \expandafter \@secondoftwo
 \fi
}%
\providecommand \@ifx [1]{%
 \ifx #1\expandafter \@firstoftwo
 \else \expandafter \@secondoftwo
 \fi
}%
\providecommand \natexlab [1]{#1}%
\providecommand \enquote  [1]{``#1''}%
\providecommand \bibnamefont  [1]{#1}%
\providecommand \bibfnamefont [1]{#1}%
\providecommand \citenamefont [1]{#1}%
\providecommand \href@noop [0]{\@secondoftwo}%
\providecommand \href [0]{\begingroup \@sanitize@url \@href}%
\providecommand \@href[1]{\@@startlink{#1}\@@href}%
\providecommand \@@href[1]{\endgroup#1\@@endlink}%
\providecommand \@sanitize@url [0]{\catcode `\\12\catcode `\$12\catcode
  `\&12\catcode `\#12\catcode `\^12\catcode `\_12\catcode `\%12\relax}%
\providecommand \@@startlink[1]{}%
\providecommand \@@endlink[0]{}%
\providecommand \url  [0]{\begingroup\@sanitize@url \@url }%
\providecommand \@url [1]{\endgroup\@href {#1}{\urlprefix }}%
\providecommand \urlprefix  [0]{URL }%
\providecommand \Eprint [0]{\href }%
\providecommand \doibase [0]{https://doi.org/}%
\providecommand \selectlanguage [0]{\@gobble}%
\providecommand \bibinfo  [0]{\@secondoftwo}%
\providecommand \bibfield  [0]{\@secondoftwo}%
\providecommand \translation [1]{[#1]}%
\providecommand \BibitemOpen [0]{}%
\providecommand \bibitemStop [0]{}%
\providecommand \bibitemNoStop [0]{.\EOS\space}%
\providecommand \EOS [0]{\spacefactor3000\relax}%
\providecommand \BibitemShut  [1]{\csname bibitem#1\endcsname}%
\let\auto@bib@innerbib\@empty
\bibitem [{\citenamefont {Nielsen}\ and\ \citenamefont
  {Chuang}(2010)}]{Nielsen2010}%
  \BibitemOpen
  \bibfield  {author} {\bibinfo {author} {\bibfnamefont {M.~A.}\ \bibnamefont
  {Nielsen}}\ and\ \bibinfo {author} {\bibfnamefont {I.~L.}\ \bibnamefont
  {Chuang}},\ }\href
  {https://www.google.com.au/books/edition/Quantum_Computation_and_Quantum_Informat/-s4DEy7o-a0C?hl=en&gbpv=0}
  {\emph {\bibinfo {title} {Quantum computation and quantum information}}},\
  \bibinfo {edition} {10th}\ ed.\ (\bibinfo  {publisher} {Cambridge University
  Press},\ \bibinfo {address} {Cambridge ; New York},\ \bibinfo {year}
  {2010})\BibitemShut {NoStop}%
\bibitem [{\citenamefont {Shor}(1997)}]{Shor1997}%
  \BibitemOpen
  \bibfield  {author} {\bibinfo {author} {\bibfnamefont {P.~W.}\ \bibnamefont
  {Shor}},\ }\href {https://doi.org/10.1137/S0097539795293172} {\bibfield
  {journal} {\bibinfo  {journal} {Siam Journal on Computing}\ }\textbf
  {\bibinfo {volume} {26}},\ \bibinfo {pages} {1484} (\bibinfo {year}
  {1997})}\BibitemShut {NoStop}%
\bibitem [{\citenamefont {Grover}(1996)}]{Grover1996}%
  \BibitemOpen
  \bibfield  {author} {\bibinfo {author} {\bibfnamefont {L.~K.}\ \bibnamefont
  {Grover}},\ }\href {https://doi.org/10.1145/237814.237866} {\bibinfo {title}
  {A fast quantum mechanical algorithm for database search}} (\bibinfo {year}
  {1996})\BibitemShut {NoStop}%
\bibitem [{\citenamefont {Wehner}\ \emph {et~al.}(2018)\citenamefont {Wehner},
  \citenamefont {Elkouss},\ and\ \citenamefont {Hanson}}]{Wehner2018}%
  \BibitemOpen
  \bibfield  {author} {\bibinfo {author} {\bibfnamefont {S.}~\bibnamefont
  {Wehner}}, \bibinfo {author} {\bibfnamefont {D.}~\bibnamefont {Elkouss}},\
  and\ \bibinfo {author} {\bibfnamefont {R.}~\bibnamefont {Hanson}},\ }\href
  {https://doi.org/10.1126/science.aam9288} {\bibfield  {journal} {\bibinfo
  {journal} {Science}\ }\textbf {\bibinfo {volume} {362}},\ \bibinfo {pages}
  {eaam9288} (\bibinfo {year} {2018})}\BibitemShut {NoStop}%
\bibitem [{\citenamefont {Childs}\ and\ \citenamefont {van
  Dam}(2010)}]{Childs2010}%
  \BibitemOpen
  \bibfield  {author} {\bibinfo {author} {\bibfnamefont {A.~M.}\ \bibnamefont
  {Childs}}\ and\ \bibinfo {author} {\bibfnamefont {W.}~\bibnamefont {van
  Dam}},\ }\href {https://doi.org/10.1103/RevModPhys.82.1} {\bibfield
  {journal} {\bibinfo  {journal} {Reviews of Modern Physics}\ }\textbf
  {\bibinfo {volume} {82}},\ \bibinfo {pages} {1} (\bibinfo {year}
  {2010})}\BibitemShut {NoStop}%
\bibitem [{\citenamefont {Divincenzo}(1995)}]{DiVincenzo1995}%
  \BibitemOpen
  \bibfield  {author} {\bibinfo {author} {\bibfnamefont {D.~P.}\ \bibnamefont
  {Divincenzo}},\ }\href {https://doi.org/10.1126/science.270.5234.255}
  {\bibfield  {journal} {\bibinfo  {journal} {Science}\ }\textbf {\bibinfo
  {volume} {270}},\ \bibinfo {pages} {255} (\bibinfo {year}
  {1995})}\BibitemShut {NoStop}%
\bibitem [{\citenamefont {Ekert}\ and\ \citenamefont
  {Jozsa}(1996)}]{Ekert1996}%
  \BibitemOpen
  \bibfield  {author} {\bibinfo {author} {\bibfnamefont {A.}~\bibnamefont
  {Ekert}}\ and\ \bibinfo {author} {\bibfnamefont {R.}~\bibnamefont {Jozsa}},\
  }\href {https://doi.org/10.1103/RevModPhys.68.733} {\bibfield  {journal}
  {\bibinfo  {journal} {Reviews of Modern Physics}\ }\textbf {\bibinfo {volume}
  {68}},\ \bibinfo {pages} {733} (\bibinfo {year} {1996})}\BibitemShut
  {NoStop}%
\bibitem [{\citenamefont {Feynman}(1982)}]{Feynman1982}%
  \BibitemOpen
  \bibfield  {author} {\bibinfo {author} {\bibfnamefont {R.~P.}\ \bibnamefont
  {Feynman}},\ }\href {https://doi.org/10.1007/Bf02650179} {\bibfield
  {journal} {\bibinfo  {journal} {International Journal of Theoretical
  Physics}\ }\textbf {\bibinfo {volume} {21}},\ \bibinfo {pages} {467}
  (\bibinfo {year} {1982})}\BibitemShut {NoStop}%
\bibitem [{\citenamefont {Jordan}(2011)}]{Jordan2011}%
  \BibitemOpen
  \bibfield  {author} {\bibinfo {author} {\bibfnamefont {S.}~\bibnamefont
  {Jordan}},\ }\href {https://quantumalgorithmzoo.org} {\bibinfo {title}
  {Quantum algorithm zoo}} (\bibinfo {year} {2011})\BibitemShut {NoStop}%
\bibitem [{\citenamefont {Montanaro}(2016)}]{Montanaro2016}%
  \BibitemOpen
  \bibfield  {author} {\bibinfo {author} {\bibfnamefont {A.}~\bibnamefont
  {Montanaro}},\ }\href {https://doi.org/10.1038/npjqi.2015.23} {\bibfield
  {journal} {\bibinfo  {journal} {Npj Quantum Information}\ }\textbf {\bibinfo
  {volume} {2}},\ \bibinfo {pages} {15023} (\bibinfo {year}
  {2016})}\BibitemShut {NoStop}%
\bibitem [{\citenamefont {Alexeev}\ \emph {et~al.}(2021)\citenamefont
  {Alexeev}, \citenamefont {Bacon}, \citenamefont {Brown}, \citenamefont
  {Calderbank}, \citenamefont {Carr}, \citenamefont {Chong}, \citenamefont
  {DeMarco}, \citenamefont {Englund}, \citenamefont {Farhi}, \citenamefont
  {Fefferman}, \citenamefont {Gorshkov}, \citenamefont {Houck}, \citenamefont
  {Kim}, \citenamefont {Kimmel}, \citenamefont {Lange}, \citenamefont {Lloyd},
  \citenamefont {Lukin}, \citenamefont {Maslov}, \citenamefont {Maunz},
  \citenamefont {Monroe}, \citenamefont {Preskill}, \citenamefont {Roetteler},
  \citenamefont {Savage},\ and\ \citenamefont {Thompson}}]{Alexeev2021}%
  \BibitemOpen
  \bibfield  {author} {\bibinfo {author} {\bibfnamefont {Y.}~\bibnamefont
  {Alexeev}}, \bibinfo {author} {\bibfnamefont {D.}~\bibnamefont {Bacon}},
  \bibinfo {author} {\bibfnamefont {K.~R.}\ \bibnamefont {Brown}}, \bibinfo
  {author} {\bibfnamefont {R.}~\bibnamefont {Calderbank}}, \bibinfo {author}
  {\bibfnamefont {L.~D.}\ \bibnamefont {Carr}}, \bibinfo {author}
  {\bibfnamefont {F.~T.}\ \bibnamefont {Chong}}, \bibinfo {author}
  {\bibfnamefont {B.}~\bibnamefont {DeMarco}}, \bibinfo {author} {\bibfnamefont
  {D.}~\bibnamefont {Englund}}, \bibinfo {author} {\bibfnamefont
  {E.}~\bibnamefont {Farhi}}, \bibinfo {author} {\bibfnamefont
  {B.}~\bibnamefont {Fefferman}}, \bibinfo {author} {\bibfnamefont {A.~V.}\
  \bibnamefont {Gorshkov}}, \bibinfo {author} {\bibfnamefont {A.}~\bibnamefont
  {Houck}}, \bibinfo {author} {\bibfnamefont {J.}~\bibnamefont {Kim}}, \bibinfo
  {author} {\bibfnamefont {S.}~\bibnamefont {Kimmel}}, \bibinfo {author}
  {\bibfnamefont {M.}~\bibnamefont {Lange}}, \bibinfo {author} {\bibfnamefont
  {S.}~\bibnamefont {Lloyd}}, \bibinfo {author} {\bibfnamefont {M.~D.}\
  \bibnamefont {Lukin}}, \bibinfo {author} {\bibfnamefont {D.}~\bibnamefont
  {Maslov}}, \bibinfo {author} {\bibfnamefont {P.}~\bibnamefont {Maunz}},
  \bibinfo {author} {\bibfnamefont {C.}~\bibnamefont {Monroe}}, \bibinfo
  {author} {\bibfnamefont {J.}~\bibnamefont {Preskill}}, \bibinfo {author}
  {\bibfnamefont {M.}~\bibnamefont {Roetteler}}, \bibinfo {author}
  {\bibfnamefont {M.~J.}\ \bibnamefont {Savage}},\ and\ \bibinfo {author}
  {\bibfnamefont {J.}~\bibnamefont {Thompson}},\ }\href
  {https://doi.org/10.1103/PRXQuantum.2.017001} {\bibfield  {journal} {\bibinfo
   {journal} {PRX Quantum}\ }\textbf {\bibinfo {volume} {2}},\ \bibinfo {pages}
  {017001} (\bibinfo {year} {2021})}\BibitemShut {NoStop}%
\bibitem [{\citenamefont {Kane}(1998)}]{Kane1998}%
  \BibitemOpen
  \bibfield  {author} {\bibinfo {author} {\bibfnamefont {B.~E.}\ \bibnamefont
  {Kane}},\ }\href {https://doi.org/10.1038/30156} {\bibfield  {journal}
  {\bibinfo  {journal} {Nature}\ }\textbf {\bibinfo {volume} {393}},\ \bibinfo
  {pages} {133} (\bibinfo {year} {1998})}\BibitemShut {NoStop}%
\bibitem [{\citenamefont {Pla}\ \emph {et~al.}(2012)\citenamefont {Pla},
  \citenamefont {Tan}, \citenamefont {Dehollain}, \citenamefont {Lim},
  \citenamefont {Morton}, \citenamefont {Jamieson}, \citenamefont {Dzurak},\
  and\ \citenamefont {Morello}}]{Pla2012}%
  \BibitemOpen
  \bibfield  {author} {\bibinfo {author} {\bibfnamefont {J.~J.}\ \bibnamefont
  {Pla}}, \bibinfo {author} {\bibfnamefont {K.~Y.}\ \bibnamefont {Tan}},
  \bibinfo {author} {\bibfnamefont {J.~P.}\ \bibnamefont {Dehollain}}, \bibinfo
  {author} {\bibfnamefont {W.~H.}\ \bibnamefont {Lim}}, \bibinfo {author}
  {\bibfnamefont {J.~J.~L.}\ \bibnamefont {Morton}}, \bibinfo {author}
  {\bibfnamefont {D.~N.}\ \bibnamefont {Jamieson}}, \bibinfo {author}
  {\bibfnamefont {A.~S.}\ \bibnamefont {Dzurak}},\ and\ \bibinfo {author}
  {\bibfnamefont {A.}~\bibnamefont {Morello}},\ }\href
  {https://doi.org/10.1038/nature11449} {\bibfield  {journal} {\bibinfo
  {journal} {Nature}\ }\textbf {\bibinfo {volume} {489}},\ \bibinfo {pages}
  {541} (\bibinfo {year} {2012})}\BibitemShut {NoStop}%
\bibitem [{\citenamefont {Fricke}\ \emph {et~al.}(2021)\citenamefont {Fricke},
  \citenamefont {Hile}, \citenamefont {Kranz}, \citenamefont {Chung},
  \citenamefont {He}, \citenamefont {Pakkiam}, \citenamefont {House},
  \citenamefont {Keizer},\ and\ \citenamefont {Simmons}}]{Fricke2021}%
  \BibitemOpen
  \bibfield  {author} {\bibinfo {author} {\bibfnamefont {L.}~\bibnamefont
  {Fricke}}, \bibinfo {author} {\bibfnamefont {S.~J.}\ \bibnamefont {Hile}},
  \bibinfo {author} {\bibfnamefont {L.}~\bibnamefont {Kranz}}, \bibinfo
  {author} {\bibfnamefont {Y.}~\bibnamefont {Chung}}, \bibinfo {author}
  {\bibfnamefont {Y.}~\bibnamefont {He}}, \bibinfo {author} {\bibfnamefont
  {P.}~\bibnamefont {Pakkiam}}, \bibinfo {author} {\bibfnamefont {M.~G.}\
  \bibnamefont {House}}, \bibinfo {author} {\bibfnamefont {J.~G.}\ \bibnamefont
  {Keizer}},\ and\ \bibinfo {author} {\bibfnamefont {M.~Y.}\ \bibnamefont
  {Simmons}},\ }\href {https://doi.org/10.1038/s41467-021-23662-3} {\bibfield
  {journal} {\bibinfo  {journal} {Nature Communications}\ }\textbf {\bibinfo
  {volume} {12}},\ \bibinfo {pages} {3323} (\bibinfo {year}
  {2021})}\BibitemShut {NoStop}%
\bibitem [{\citenamefont {Philips}\ \emph {et~al.}(2022)\citenamefont
  {Philips}, \citenamefont {Amitonov}, \citenamefont {de~Snoo}, \citenamefont
  {Russ}, \citenamefont {Kalhor}, \citenamefont {Volk}, \citenamefont {Lawrie},
  \citenamefont {Brousse}, \citenamefont {Tryputen}, \citenamefont {Wuetz},
  \citenamefont {Sammak}, \citenamefont {Veldhorst}, \citenamefont
  {Scappucci},\ and\ \citenamefont {Vandersypen}}]{Philips2022}%
  \BibitemOpen
  \bibfield  {author} {\bibinfo {author} {\bibfnamefont {S.~G.~J.}\
  \bibnamefont {Philips}}, \bibinfo {author} {\bibfnamefont {S.~V.}\
  \bibnamefont {Amitonov}}, \bibinfo {author} {\bibfnamefont {S.~L.}\
  \bibnamefont {de~Snoo}}, \bibinfo {author} {\bibfnamefont {M.}~\bibnamefont
  {Russ}}, \bibinfo {author} {\bibfnamefont {N.}~\bibnamefont {Kalhor}},
  \bibinfo {author} {\bibfnamefont {C.}~\bibnamefont {Volk}}, \bibinfo {author}
  {\bibfnamefont {W.~I.~L.}\ \bibnamefont {Lawrie}}, \bibinfo {author}
  {\bibfnamefont {D.}~\bibnamefont {Brousse}}, \bibinfo {author} {\bibfnamefont
  {L.}~\bibnamefont {Tryputen}}, \bibinfo {author} {\bibfnamefont {B.~P.}\
  \bibnamefont {Wuetz}}, \bibinfo {author} {\bibfnamefont {A.}~\bibnamefont
  {Sammak}}, \bibinfo {author} {\bibfnamefont {M.}~\bibnamefont {Veldhorst}},
  \bibinfo {author} {\bibfnamefont {G.}~\bibnamefont {Scappucci}},\ and\
  \bibinfo {author} {\bibfnamefont {L.~M.~K.}\ \bibnamefont {Vandersypen}},\
  }\href {https://doi.org/10.1038/s41586-022-05117-x} {\bibfield  {journal}
  {\bibinfo  {journal} {Nature}\ }\textbf {\bibinfo {volume} {609}},\ \bibinfo
  {pages} {919} (\bibinfo {year} {2022})}\BibitemShut {NoStop}%
\bibitem [{\citenamefont {Burkard}\ \emph {et~al.}(2021)\citenamefont
  {Burkard}, \citenamefont {Ladd}, \citenamefont {Nichol}, \citenamefont
  {Pan},\ and\ \citenamefont {Petta}}]{Burkard2021}%
  \BibitemOpen
  \bibfield  {author} {\bibinfo {author} {\bibfnamefont {G.}~\bibnamefont
  {Burkard}}, \bibinfo {author} {\bibfnamefont {T.~D.}\ \bibnamefont {Ladd}},
  \bibinfo {author} {\bibfnamefont {J.~M.}\ \bibnamefont {Nichol}}, \bibinfo
  {author} {\bibfnamefont {A.}~\bibnamefont {Pan}},\ and\ \bibinfo {author}
  {\bibfnamefont {J.~R.}\ \bibnamefont {Petta}},\ }\href
  {https://ui.adsabs.harvard.edu/abs/2021arXiv211208863B} {\bibinfo {title}
  {Semiconductor spin qubits}} (\bibinfo {year} {2021})\BibitemShut {NoStop}%
\bibitem [{\citenamefont {Tyryshkin}\ \emph {et~al.}(2012)\citenamefont
  {Tyryshkin}, \citenamefont {Tojo}, \citenamefont {Morton}, \citenamefont
  {Riemann}, \citenamefont {Abrosimov}, \citenamefont {Becker}, \citenamefont
  {Pohl}, \citenamefont {Schenkel}, \citenamefont {Thewalt}, \citenamefont
  {Itoh},\ and\ \citenamefont {Lyon}}]{Tyryshkin2012}%
  \BibitemOpen
  \bibfield  {author} {\bibinfo {author} {\bibfnamefont {A.~M.}\ \bibnamefont
  {Tyryshkin}}, \bibinfo {author} {\bibfnamefont {S.}~\bibnamefont {Tojo}},
  \bibinfo {author} {\bibfnamefont {J.~J.~L.}\ \bibnamefont {Morton}}, \bibinfo
  {author} {\bibfnamefont {H.}~\bibnamefont {Riemann}}, \bibinfo {author}
  {\bibfnamefont {N.~V.}\ \bibnamefont {Abrosimov}}, \bibinfo {author}
  {\bibfnamefont {P.}~\bibnamefont {Becker}}, \bibinfo {author} {\bibfnamefont
  {H.-J.}\ \bibnamefont {Pohl}}, \bibinfo {author} {\bibfnamefont
  {T.}~\bibnamefont {Schenkel}}, \bibinfo {author} {\bibfnamefont {M.~L.~W.}\
  \bibnamefont {Thewalt}}, \bibinfo {author} {\bibfnamefont {K.~M.}\
  \bibnamefont {Itoh}},\ and\ \bibinfo {author} {\bibfnamefont {S.~A.}\
  \bibnamefont {Lyon}},\ }\href {https://doi.org/10.1038/nmat3182} {\bibfield
  {journal} {\bibinfo  {journal} {Nature Materials}\ }\textbf {\bibinfo
  {volume} {11}},\ \bibinfo {pages} {143} (\bibinfo {year} {2012})}\BibitemShut
  {NoStop}%
\bibitem [{\citenamefont {Veldhorst}\ \emph {et~al.}(2014)\citenamefont
  {Veldhorst}, \citenamefont {Hwang}, \citenamefont {Yang}, \citenamefont
  {Leenstra}, \citenamefont {de~Ronde}, \citenamefont {Dehollain},
  \citenamefont {Muhonen}, \citenamefont {Hudson}, \citenamefont {Itoh},
  \citenamefont {Morello},\ and\ \citenamefont {Dzurak}}]{Veldhorst2014}%
  \BibitemOpen
  \bibfield  {author} {\bibinfo {author} {\bibfnamefont {M.}~\bibnamefont
  {Veldhorst}}, \bibinfo {author} {\bibfnamefont {J.~C.~C.}\ \bibnamefont
  {Hwang}}, \bibinfo {author} {\bibfnamefont {C.~H.}\ \bibnamefont {Yang}},
  \bibinfo {author} {\bibfnamefont {A.~W.}\ \bibnamefont {Leenstra}}, \bibinfo
  {author} {\bibfnamefont {B.}~\bibnamefont {de~Ronde}}, \bibinfo {author}
  {\bibfnamefont {J.~P.}\ \bibnamefont {Dehollain}}, \bibinfo {author}
  {\bibfnamefont {J.~T.}\ \bibnamefont {Muhonen}}, \bibinfo {author}
  {\bibfnamefont {F.~E.}\ \bibnamefont {Hudson}}, \bibinfo {author}
  {\bibfnamefont {K.~M.}\ \bibnamefont {Itoh}}, \bibinfo {author}
  {\bibfnamefont {A.}~\bibnamefont {Morello}},\ and\ \bibinfo {author}
  {\bibfnamefont {A.~S.}\ \bibnamefont {Dzurak}},\ }\href
  {https://doi.org/10.1038/nnano.2014.216} {\bibfield  {journal} {\bibinfo
  {journal} {Nature Nanotechnology}\ }\textbf {\bibinfo {volume} {9}},\
  \bibinfo {pages} {981} (\bibinfo {year} {2014})}\BibitemShut {NoStop}%
\bibitem [{\citenamefont {Takeda}\ \emph {et~al.}(2022)\citenamefont {Takeda},
  \citenamefont {Noiri}, \citenamefont {Nakajima}, \citenamefont {Kobayashi},\
  and\ \citenamefont {Tarucha}}]{Takeda2022}%
  \BibitemOpen
  \bibfield  {author} {\bibinfo {author} {\bibfnamefont {K.}~\bibnamefont
  {Takeda}}, \bibinfo {author} {\bibfnamefont {A.}~\bibnamefont {Noiri}},
  \bibinfo {author} {\bibfnamefont {T.}~\bibnamefont {Nakajima}}, \bibinfo
  {author} {\bibfnamefont {T.}~\bibnamefont {Kobayashi}},\ and\ \bibinfo
  {author} {\bibfnamefont {S.}~\bibnamefont {Tarucha}},\ }\href
  {https://doi.org/10.1038/s41586-022-04986-6} {\bibfield  {journal} {\bibinfo
  {journal} {Nature}\ }\textbf {\bibinfo {volume} {608}},\ \bibinfo {pages}
  {682} (\bibinfo {year} {2022})}\BibitemShut {NoStop}%
\bibitem [{\citenamefont {Veldhorst}\ \emph {et~al.}(2015)\citenamefont
  {Veldhorst}, \citenamefont {Yang}, \citenamefont {Hwang}, \citenamefont
  {Huang}, \citenamefont {Dehollain}, \citenamefont {Muhonen}, \citenamefont
  {Simmons}, \citenamefont {Laucht}, \citenamefont {Hudson}, \citenamefont
  {Itoh}, \citenamefont {Morello},\ and\ \citenamefont
  {Dzurak}}]{Veldhorst2015}%
  \BibitemOpen
  \bibfield  {author} {\bibinfo {author} {\bibfnamefont {M.}~\bibnamefont
  {Veldhorst}}, \bibinfo {author} {\bibfnamefont {C.~H.}\ \bibnamefont {Yang}},
  \bibinfo {author} {\bibfnamefont {J.~C.~C.}\ \bibnamefont {Hwang}}, \bibinfo
  {author} {\bibfnamefont {W.}~\bibnamefont {Huang}}, \bibinfo {author}
  {\bibfnamefont {J.~P.}\ \bibnamefont {Dehollain}}, \bibinfo {author}
  {\bibfnamefont {J.~T.}\ \bibnamefont {Muhonen}}, \bibinfo {author}
  {\bibfnamefont {S.}~\bibnamefont {Simmons}}, \bibinfo {author} {\bibfnamefont
  {A.}~\bibnamefont {Laucht}}, \bibinfo {author} {\bibfnamefont {F.~E.}\
  \bibnamefont {Hudson}}, \bibinfo {author} {\bibfnamefont {K.~M.}\
  \bibnamefont {Itoh}}, \bibinfo {author} {\bibfnamefont {A.}~\bibnamefont
  {Morello}},\ and\ \bibinfo {author} {\bibfnamefont {A.~S.}\ \bibnamefont
  {Dzurak}},\ }\href {https://doi.org/10.1038/nature15263} {\bibfield
  {journal} {\bibinfo  {journal} {Nature}\ }\textbf {\bibinfo {volume} {526}},\
  \bibinfo {pages} {410} (\bibinfo {year} {2015})}\BibitemShut {NoStop}%
\bibitem [{\citenamefont {Xue}\ \emph {et~al.}(2022)\citenamefont {Xue},
  \citenamefont {Russ}, \citenamefont {Samkharadze}, \citenamefont {Undseth},
  \citenamefont {Sammak}, \citenamefont {Scappucci},\ and\ \citenamefont
  {Vandersypen}}]{Xue2022}%
  \BibitemOpen
  \bibfield  {author} {\bibinfo {author} {\bibfnamefont {X.}~\bibnamefont
  {Xue}}, \bibinfo {author} {\bibfnamefont {M.}~\bibnamefont {Russ}}, \bibinfo
  {author} {\bibfnamefont {N.}~\bibnamefont {Samkharadze}}, \bibinfo {author}
  {\bibfnamefont {B.}~\bibnamefont {Undseth}}, \bibinfo {author} {\bibfnamefont
  {A.}~\bibnamefont {Sammak}}, \bibinfo {author} {\bibfnamefont
  {G.}~\bibnamefont {Scappucci}},\ and\ \bibinfo {author} {\bibfnamefont
  {L.~M.~K.}\ \bibnamefont {Vandersypen}},\ }\href
  {https://doi.org/10.1038/s41586-021-04273-w} {\bibfield  {journal} {\bibinfo
  {journal} {Nature}\ }\textbf {\bibinfo {volume} {601}},\ \bibinfo {pages}
  {343} (\bibinfo {year} {2022})}\BibitemShut {NoStop}%
\bibitem [{\citenamefont {Yoneda}\ \emph {et~al.}(2018)\citenamefont {Yoneda},
  \citenamefont {Takeda}, \citenamefont {Otsuka}, \citenamefont {Nakajima},
  \citenamefont {Delbecq}, \citenamefont {Allison}, \citenamefont {Honda},
  \citenamefont {Kodera}, \citenamefont {Oda}, \citenamefont {Hoshi},
  \citenamefont {Usami}, \citenamefont {Itoh},\ and\ \citenamefont
  {Tarucha}}]{Yoneda2018}%
  \BibitemOpen
  \bibfield  {author} {\bibinfo {author} {\bibfnamefont {J.}~\bibnamefont
  {Yoneda}}, \bibinfo {author} {\bibfnamefont {K.}~\bibnamefont {Takeda}},
  \bibinfo {author} {\bibfnamefont {T.}~\bibnamefont {Otsuka}}, \bibinfo
  {author} {\bibfnamefont {T.}~\bibnamefont {Nakajima}}, \bibinfo {author}
  {\bibfnamefont {M.~R.}\ \bibnamefont {Delbecq}}, \bibinfo {author}
  {\bibfnamefont {G.}~\bibnamefont {Allison}}, \bibinfo {author} {\bibfnamefont
  {T.}~\bibnamefont {Honda}}, \bibinfo {author} {\bibfnamefont
  {T.}~\bibnamefont {Kodera}}, \bibinfo {author} {\bibfnamefont
  {S.}~\bibnamefont {Oda}}, \bibinfo {author} {\bibfnamefont {Y.}~\bibnamefont
  {Hoshi}}, \bibinfo {author} {\bibfnamefont {N.}~\bibnamefont {Usami}},
  \bibinfo {author} {\bibfnamefont {K.~M.}\ \bibnamefont {Itoh}},\ and\
  \bibinfo {author} {\bibfnamefont {S.}~\bibnamefont {Tarucha}},\ }\href
  {https://doi.org/10.1038/s41565-017-0014-x} {\bibfield  {journal} {\bibinfo
  {journal} {Nature Nanotechnology}\ }\textbf {\bibinfo {volume} {13}},\
  \bibinfo {pages} {102} (\bibinfo {year} {2018})}\BibitemShut {NoStop}%
\bibitem [{\citenamefont {Bergeron}\ \emph {et~al.}(2020)\citenamefont
  {Bergeron}, \citenamefont {Chartrand}, \citenamefont {Kurkjian},
  \citenamefont {Morse}, \citenamefont {Riemann}, \citenamefont {Abrosimov},
  \citenamefont {Becker}, \citenamefont {Pohl}, \citenamefont {Thewalt},\ and\
  \citenamefont {Simmons}}]{Bergeron2020}%
  \BibitemOpen
  \bibfield  {author} {\bibinfo {author} {\bibfnamefont {L.}~\bibnamefont
  {Bergeron}}, \bibinfo {author} {\bibfnamefont {C.}~\bibnamefont {Chartrand}},
  \bibinfo {author} {\bibfnamefont {A.~T.~K.}\ \bibnamefont {Kurkjian}},
  \bibinfo {author} {\bibfnamefont {K.~J.}\ \bibnamefont {Morse}}, \bibinfo
  {author} {\bibfnamefont {H.}~\bibnamefont {Riemann}}, \bibinfo {author}
  {\bibfnamefont {N.~V.}\ \bibnamefont {Abrosimov}}, \bibinfo {author}
  {\bibfnamefont {P.}~\bibnamefont {Becker}}, \bibinfo {author} {\bibfnamefont
  {H.~J.}\ \bibnamefont {Pohl}}, \bibinfo {author} {\bibfnamefont {M.~L.~W.}\
  \bibnamefont {Thewalt}},\ and\ \bibinfo {author} {\bibfnamefont
  {S.}~\bibnamefont {Simmons}},\ }\href
  {https://doi.org/10.1103/PRXQuantum.1.020301} {\bibfield  {journal} {\bibinfo
   {journal} {PRX Quantum}\ }\textbf {\bibinfo {volume} {1}},\ \bibinfo {pages}
  {020301} (\bibinfo {year} {2020})}\BibitemShut {NoStop}%
\bibitem [{\citenamefont {El-Kareh}(1995)}]{ElKareh1995}%
  \BibitemOpen
  \bibfield  {author} {\bibinfo {author} {\bibfnamefont {B.}~\bibnamefont
  {El-Kareh}},\ }\href
  {https://www.google.com.au/books/edition/Fundamentals_of_Semiconductor_Processing/K0jVBwAAQBAJ?hl=en&gbpv=0}
  {\emph {\bibinfo {title} {Fundamentals of semiconductor processing
  technologies}}}\ (\bibinfo  {publisher} {Kluwer Academic Publishers},\
  \bibinfo {address} {Boston},\ \bibinfo {year} {1995})\ pp.\ \bibinfo {pages}
  {xii, 599 p.}\BibitemShut {Stop}%
\bibitem [{\citenamefont {Li}\ \emph {et~al.}(2018)\citenamefont {Li},
  \citenamefont {Petit}, \citenamefont {Franke}, \citenamefont {Dehollain},
  \citenamefont {Helsen}, \citenamefont {Steudtner}, \citenamefont {Thomas},
  \citenamefont {Yoscovits}, \citenamefont {Singh}, \citenamefont {Wehner},
  \citenamefont {Vandersypen}, \citenamefont {Clarke},\ and\ \citenamefont
  {Veldhorst}}]{Li2018}%
  \BibitemOpen
  \bibfield  {author} {\bibinfo {author} {\bibfnamefont {R.}~\bibnamefont
  {Li}}, \bibinfo {author} {\bibfnamefont {L.}~\bibnamefont {Petit}}, \bibinfo
  {author} {\bibfnamefont {D.~P.}\ \bibnamefont {Franke}}, \bibinfo {author}
  {\bibfnamefont {J.~P.}\ \bibnamefont {Dehollain}}, \bibinfo {author}
  {\bibfnamefont {J.}~\bibnamefont {Helsen}}, \bibinfo {author} {\bibfnamefont
  {M.}~\bibnamefont {Steudtner}}, \bibinfo {author} {\bibfnamefont {N.~K.}\
  \bibnamefont {Thomas}}, \bibinfo {author} {\bibfnamefont {Z.~R.}\
  \bibnamefont {Yoscovits}}, \bibinfo {author} {\bibfnamefont {K.~J.}\
  \bibnamefont {Singh}}, \bibinfo {author} {\bibfnamefont {S.}~\bibnamefont
  {Wehner}}, \bibinfo {author} {\bibfnamefont {L.~M.~K.}\ \bibnamefont
  {Vandersypen}}, \bibinfo {author} {\bibfnamefont {J.~S.}\ \bibnamefont
  {Clarke}},\ and\ \bibinfo {author} {\bibfnamefont {M.}~\bibnamefont
  {Veldhorst}},\ }\href {https://doi.org/10.1126/sciadv.aar3960} {\bibfield
  {journal} {\bibinfo  {journal} {Science Advances}\ }\textbf {\bibinfo
  {volume} {4}},\ \bibinfo {pages} {eaar3960} (\bibinfo {year}
  {2018})}\BibitemShut {NoStop}%
\bibitem [{\citenamefont {Jones}\ \emph {et~al.}(2018)\citenamefont {Jones},
  \citenamefont {Fogarty}, \citenamefont {Morello}, \citenamefont {Gyure},
  \citenamefont {Dzurak},\ and\ \citenamefont {Ladd}}]{Jones2018}%
  \BibitemOpen
  \bibfield  {author} {\bibinfo {author} {\bibfnamefont {C.}~\bibnamefont
  {Jones}}, \bibinfo {author} {\bibfnamefont {M.~A.}\ \bibnamefont {Fogarty}},
  \bibinfo {author} {\bibfnamefont {A.}~\bibnamefont {Morello}}, \bibinfo
  {author} {\bibfnamefont {M.~F.}\ \bibnamefont {Gyure}}, \bibinfo {author}
  {\bibfnamefont {A.~S.}\ \bibnamefont {Dzurak}},\ and\ \bibinfo {author}
  {\bibfnamefont {T.~D.}\ \bibnamefont {Ladd}},\ }\href
  {https://doi.org/10.1103/PhysRevX.8.021058} {\bibfield  {journal} {\bibinfo
  {journal} {Physical Review X}\ }\textbf {\bibinfo {volume} {8}},\ \bibinfo
  {pages} {021058} (\bibinfo {year} {2018})}\BibitemShut {NoStop}%
\bibitem [{\citenamefont {Zajac}\ \emph {et~al.}(2016)\citenamefont {Zajac},
  \citenamefont {Hazard}, \citenamefont {Mi}, \citenamefont {Nielsen},\ and\
  \citenamefont {Petta}}]{Zajac2016}%
  \BibitemOpen
  \bibfield  {author} {\bibinfo {author} {\bibfnamefont {D.~M.}\ \bibnamefont
  {Zajac}}, \bibinfo {author} {\bibfnamefont {T.~M.}\ \bibnamefont {Hazard}},
  \bibinfo {author} {\bibfnamefont {X.}~\bibnamefont {Mi}}, \bibinfo {author}
  {\bibfnamefont {E.}~\bibnamefont {Nielsen}},\ and\ \bibinfo {author}
  {\bibfnamefont {J.~R.}\ \bibnamefont {Petta}},\ }\href
  {https://doi.org/10.1103/PhysRevApplied.6.054013} {\bibfield  {journal}
  {\bibinfo  {journal} {Physical Review Applied}\ }\textbf {\bibinfo {volume}
  {6}},\ \bibinfo {pages} {054013} (\bibinfo {year} {2016})}\BibitemShut
  {NoStop}%
\bibitem [{\citenamefont {Zwerver}\ \emph {et~al.}(2022)\citenamefont
  {Zwerver}, \citenamefont {Krahenmann}, \citenamefont {Watson}, \citenamefont
  {Lampert}, \citenamefont {George}, \citenamefont {Pillarisetty},
  \citenamefont {Bojarski}, \citenamefont {Amin}, \citenamefont {Amitonov},
  \citenamefont {Boter}, \citenamefont {Caudillo}, \citenamefont
  {Corras-Serrano}, \citenamefont {Dehollain}, \citenamefont {Droulers},
  \citenamefont {Henry}, \citenamefont {Kotlyar}, \citenamefont {Lodari},
  \citenamefont {Luthi}, \citenamefont {Michalak}, \citenamefont {Mueller},
  \citenamefont {Neyens}, \citenamefont {Roberts}, \citenamefont {Samkharadze},
  \citenamefont {Zheng}, \citenamefont {Zietz}, \citenamefont {Scappucci},
  \citenamefont {Veldhorst}, \citenamefont {Vandersypen},\ and\ \citenamefont
  {Clarke}}]{Zwerver2022}%
  \BibitemOpen
  \bibfield  {author} {\bibinfo {author} {\bibfnamefont {A.~M.~J.}\
  \bibnamefont {Zwerver}}, \bibinfo {author} {\bibfnamefont {T.}~\bibnamefont
  {Krahenmann}}, \bibinfo {author} {\bibfnamefont {T.~F.}\ \bibnamefont
  {Watson}}, \bibinfo {author} {\bibfnamefont {L.}~\bibnamefont {Lampert}},
  \bibinfo {author} {\bibfnamefont {H.~C.}\ \bibnamefont {George}}, \bibinfo
  {author} {\bibfnamefont {R.}~\bibnamefont {Pillarisetty}}, \bibinfo {author}
  {\bibfnamefont {S.~A.}\ \bibnamefont {Bojarski}}, \bibinfo {author}
  {\bibfnamefont {P.}~\bibnamefont {Amin}}, \bibinfo {author} {\bibfnamefont
  {S.~V.}\ \bibnamefont {Amitonov}}, \bibinfo {author} {\bibfnamefont {J.~M.}\
  \bibnamefont {Boter}}, \bibinfo {author} {\bibfnamefont {R.}~\bibnamefont
  {Caudillo}}, \bibinfo {author} {\bibfnamefont {D.}~\bibnamefont
  {Corras-Serrano}}, \bibinfo {author} {\bibfnamefont {J.~P.}\ \bibnamefont
  {Dehollain}}, \bibinfo {author} {\bibfnamefont {G.}~\bibnamefont {Droulers}},
  \bibinfo {author} {\bibfnamefont {E.~M.}\ \bibnamefont {Henry}}, \bibinfo
  {author} {\bibfnamefont {R.}~\bibnamefont {Kotlyar}}, \bibinfo {author}
  {\bibfnamefont {M.}~\bibnamefont {Lodari}}, \bibinfo {author} {\bibfnamefont
  {F.}~\bibnamefont {Luthi}}, \bibinfo {author} {\bibfnamefont {D.~J.}\
  \bibnamefont {Michalak}}, \bibinfo {author} {\bibfnamefont {B.~K.}\
  \bibnamefont {Mueller}}, \bibinfo {author} {\bibfnamefont {S.}~\bibnamefont
  {Neyens}}, \bibinfo {author} {\bibfnamefont {J.}~\bibnamefont {Roberts}},
  \bibinfo {author} {\bibfnamefont {N.}~\bibnamefont {Samkharadze}}, \bibinfo
  {author} {\bibfnamefont {G.}~\bibnamefont {Zheng}}, \bibinfo {author}
  {\bibfnamefont {O.~K.}\ \bibnamefont {Zietz}}, \bibinfo {author}
  {\bibfnamefont {G.}~\bibnamefont {Scappucci}}, \bibinfo {author}
  {\bibfnamefont {M.}~\bibnamefont {Veldhorst}}, \bibinfo {author}
  {\bibfnamefont {L.~M.~K.}\ \bibnamefont {Vandersypen}},\ and\ \bibinfo
  {author} {\bibfnamefont {J.~S.}\ \bibnamefont {Clarke}},\ }\href
  {https://doi.org/10.1038/s41928-022-00727-9} {\bibfield  {journal} {\bibinfo
  {journal} {Nature Electronics}\ }\textbf {\bibinfo {volume} {5}},\ \bibinfo
  {pages} {184} (\bibinfo {year} {2022})}\BibitemShut {NoStop}%
\bibitem [{\citenamefont {Vandersypen}\ \emph {et~al.}(2017)\citenamefont
  {Vandersypen}, \citenamefont {Bluhm}, \citenamefont {Clarke}, \citenamefont
  {Dzurak}, \citenamefont {Ishihara}, \citenamefont {Morello}, \citenamefont
  {Reilly}, \citenamefont {Schreiber},\ and\ \citenamefont
  {Veldhorst}}]{Vandersypen2017}%
  \BibitemOpen
  \bibfield  {author} {\bibinfo {author} {\bibfnamefont {L.~M.~K.}\
  \bibnamefont {Vandersypen}}, \bibinfo {author} {\bibfnamefont
  {H.}~\bibnamefont {Bluhm}}, \bibinfo {author} {\bibfnamefont {J.~S.}\
  \bibnamefont {Clarke}}, \bibinfo {author} {\bibfnamefont {A.~S.}\
  \bibnamefont {Dzurak}}, \bibinfo {author} {\bibfnamefont {R.}~\bibnamefont
  {Ishihara}}, \bibinfo {author} {\bibfnamefont {A.}~\bibnamefont {Morello}},
  \bibinfo {author} {\bibfnamefont {D.~J.}\ \bibnamefont {Reilly}}, \bibinfo
  {author} {\bibfnamefont {L.~R.}\ \bibnamefont {Schreiber}},\ and\ \bibinfo
  {author} {\bibfnamefont {M.}~\bibnamefont {Veldhorst}},\ }\href
  {https://doi.org/10.1038/s41534-017-0038-y} {\bibfield  {journal} {\bibinfo
  {journal} {npj Quantum Information}\ }\textbf {\bibinfo {volume} {3}},\
  \bibinfo {pages} {34} (\bibinfo {year} {2017})}\BibitemShut {NoStop}%
\bibitem [{\citenamefont {Borjans}\ \emph {et~al.}(2020)\citenamefont
  {Borjans}, \citenamefont {Croot}, \citenamefont {Mi}, \citenamefont
  {Gullans},\ and\ \citenamefont {Petta}}]{Borjans2020}%
  \BibitemOpen
  \bibfield  {author} {\bibinfo {author} {\bibfnamefont {F.}~\bibnamefont
  {Borjans}}, \bibinfo {author} {\bibfnamefont {X.~G.}\ \bibnamefont {Croot}},
  \bibinfo {author} {\bibfnamefont {X.}~\bibnamefont {Mi}}, \bibinfo {author}
  {\bibfnamefont {M.~J.}\ \bibnamefont {Gullans}},\ and\ \bibinfo {author}
  {\bibfnamefont {J.~R.}\ \bibnamefont {Petta}},\ }\href
  {https://doi.org/10.1038/s41586-019-1867-y} {\bibfield  {journal} {\bibinfo
  {journal} {Nature}\ }\textbf {\bibinfo {volume} {577}},\ \bibinfo {pages}
  {195} (\bibinfo {year} {2020})}\BibitemShut {NoStop}%
\bibitem [{\citenamefont {Yan}\ \emph {et~al.}(2021)\citenamefont {Yan},
  \citenamefont {Gitt}, \citenamefont {Lin}, \citenamefont {Witt},
  \citenamefont {Abdolahi}, \citenamefont {Afifi}, \citenamefont {Azem},
  \citenamefont {Darcie}, \citenamefont {Wu}, \citenamefont {Awan},
  \citenamefont {Mitchell}, \citenamefont {Pfenning}, \citenamefont
  {Chrostowski},\ and\ \citenamefont {Young}}]{Yan2021}%
  \BibitemOpen
  \bibfield  {author} {\bibinfo {author} {\bibfnamefont {X.~R.}\ \bibnamefont
  {Yan}}, \bibinfo {author} {\bibfnamefont {S.}~\bibnamefont {Gitt}}, \bibinfo
  {author} {\bibfnamefont {B.}~\bibnamefont {Lin}}, \bibinfo {author}
  {\bibfnamefont {D.}~\bibnamefont {Witt}}, \bibinfo {author} {\bibfnamefont
  {M.}~\bibnamefont {Abdolahi}}, \bibinfo {author} {\bibfnamefont
  {A.}~\bibnamefont {Afifi}}, \bibinfo {author} {\bibfnamefont
  {A.}~\bibnamefont {Azem}}, \bibinfo {author} {\bibfnamefont {A.}~\bibnamefont
  {Darcie}}, \bibinfo {author} {\bibfnamefont {J.~D.}\ \bibnamefont {Wu}},
  \bibinfo {author} {\bibfnamefont {K.}~\bibnamefont {Awan}}, \bibinfo {author}
  {\bibfnamefont {M.}~\bibnamefont {Mitchell}}, \bibinfo {author}
  {\bibfnamefont {A.}~\bibnamefont {Pfenning}}, \bibinfo {author}
  {\bibfnamefont {L.}~\bibnamefont {Chrostowski}},\ and\ \bibinfo {author}
  {\bibfnamefont {J.~F.}\ \bibnamefont {Young}},\ }\href
  {https://doi.org/10.1063/5.0049372} {\bibfield  {journal} {\bibinfo
  {journal} {APL Photonics}\ }\textbf {\bibinfo {volume} {6}},\ \bibinfo
  {pages} {070901} (\bibinfo {year} {2021})}\BibitemShut {NoStop}%
\bibitem [{\citenamefont {Harvey-Collard}\ \emph {et~al.}(2022)\citenamefont
  {Harvey-Collard}, \citenamefont {Dijkema}, \citenamefont {Zheng},
  \citenamefont {Sammak}, \citenamefont {Scappucci},\ and\ \citenamefont
  {Vandersypen}}]{Harvey2022}%
  \BibitemOpen
  \bibfield  {author} {\bibinfo {author} {\bibfnamefont {P.}~\bibnamefont
  {Harvey-Collard}}, \bibinfo {author} {\bibfnamefont {J.}~\bibnamefont
  {Dijkema}}, \bibinfo {author} {\bibfnamefont {G.~J.}\ \bibnamefont {Zheng}},
  \bibinfo {author} {\bibfnamefont {A.}~\bibnamefont {Sammak}}, \bibinfo
  {author} {\bibfnamefont {G.}~\bibnamefont {Scappucci}},\ and\ \bibinfo
  {author} {\bibfnamefont {L.~M.~K.}\ \bibnamefont {Vandersypen}},\ }\href
  {https://doi.org/10.1103/PhysRevX.12.021026} {\bibfield  {journal} {\bibinfo
  {journal} {Physical Review X}\ }\textbf {\bibinfo {volume} {12}},\ \bibinfo
  {pages} {021026} (\bibinfo {year} {2022})}\BibitemShut {NoStop}%
\bibitem [{\citenamefont {Kimble}(2008)}]{Kimble2008}%
  \BibitemOpen
  \bibfield  {author} {\bibinfo {author} {\bibfnamefont {H.~J.}\ \bibnamefont
  {Kimble}},\ }\href {https://doi.org/10.1038/nature07127} {\bibfield
  {journal} {\bibinfo  {journal} {Nature}\ }\textbf {\bibinfo {volume} {453}},\
  \bibinfo {pages} {1023} (\bibinfo {year} {2008})}\BibitemShut {NoStop}%
\bibitem [{\citenamefont {Borregaard}\ \emph {et~al.}(2019)\citenamefont
  {Borregaard}, \citenamefont {Sorensen},\ and\ \citenamefont
  {Lodahl}}]{Borregaard2019}%
  \BibitemOpen
  \bibfield  {author} {\bibinfo {author} {\bibfnamefont {J.}~\bibnamefont
  {Borregaard}}, \bibinfo {author} {\bibfnamefont {A.~S.}\ \bibnamefont
  {Sorensen}},\ and\ \bibinfo {author} {\bibfnamefont {P.}~\bibnamefont
  {Lodahl}},\ }\href {https://doi.org/10.1002/qute.201800091} {\bibfield
  {journal} {\bibinfo  {journal} {Advanced Quantum Technologies}\ }\textbf
  {\bibinfo {volume} {2}},\ \bibinfo {pages} {1800091} (\bibinfo {year}
  {2019})}\BibitemShut {NoStop}%
\bibitem [{\citenamefont {Zhu}\ \emph {et~al.}(2020)\citenamefont {Zhu},
  \citenamefont {Tu}, \citenamefont {Guo}, \citenamefont {Zhou}, \citenamefont
  {Guo},\ and\ \citenamefont {Li}}]{Zhu2020}%
  \BibitemOpen
  \bibfield  {author} {\bibinfo {author} {\bibfnamefont {X.~Y.}\ \bibnamefont
  {Zhu}}, \bibinfo {author} {\bibfnamefont {T.}~\bibnamefont {Tu}}, \bibinfo
  {author} {\bibfnamefont {A.~L.}\ \bibnamefont {Guo}}, \bibinfo {author}
  {\bibfnamefont {Z.~Q.}\ \bibnamefont {Zhou}}, \bibinfo {author}
  {\bibfnamefont {G.~C.}\ \bibnamefont {Guo}},\ and\ \bibinfo {author}
  {\bibfnamefont {C.~F.}\ \bibnamefont {Li}},\ }\href
  {https://doi.org/10.1038/s41598-020-61976-2} {\bibfield  {journal} {\bibinfo
  {journal} {Scientific Reports}\ }\textbf {\bibinfo {volume} {10}},\ \bibinfo
  {pages} {5063} (\bibinfo {year} {2020})}\BibitemShut {NoStop}%
\bibitem [{\citenamefont {Singal}(2017)}]{Singal2017}%
  \BibitemOpen
  \bibfield  {author} {\bibinfo {author} {\bibfnamefont {T.}~\bibnamefont
  {Singal}},\ }\href {https://books.google.com.au/books?id=_JheDwAAQBAJ} {\emph
  {\bibinfo {title} {Optical Fiber Communications: Principles and
  Applications}}}\ (\bibinfo  {publisher} {Cambridge University Press},\
  \bibinfo {year} {2017})\BibitemShut {NoStop}%
\bibitem [{\citenamefont {Panuski}\ \emph {et~al.}(2020)\citenamefont
  {Panuski}, \citenamefont {Englund},\ and\ \citenamefont
  {Hamerly}}]{Panuski2020}%
  \BibitemOpen
  \bibfield  {author} {\bibinfo {author} {\bibfnamefont {C.}~\bibnamefont
  {Panuski}}, \bibinfo {author} {\bibfnamefont {D.}~\bibnamefont {Englund}},\
  and\ \bibinfo {author} {\bibfnamefont {R.}~\bibnamefont {Hamerly}},\ }\href
  {https://doi.org/10.1103/PhysRevX.10.041046} {\bibfield  {journal} {\bibinfo
  {journal} {Physical Review X}\ }\textbf {\bibinfo {volume} {10}},\ \bibinfo
  {pages} {041046} (\bibinfo {year} {2020})}\BibitemShut {NoStop}%
\bibitem [{\citenamefont {Liu}\ \emph {et~al.}(2021)\citenamefont {Liu},
  \citenamefont {Huang}, \citenamefont {Wang}, \citenamefont {He},
  \citenamefont {Raja}, \citenamefont {Liu}, \citenamefont {Engelsen},\ and\
  \citenamefont {Kippenberg}}]{Liu2021}%
  \BibitemOpen
  \bibfield  {author} {\bibinfo {author} {\bibfnamefont {J.~Q.}\ \bibnamefont
  {Liu}}, \bibinfo {author} {\bibfnamefont {G.~H.}\ \bibnamefont {Huang}},
  \bibinfo {author} {\bibfnamefont {R.~N.}\ \bibnamefont {Wang}}, \bibinfo
  {author} {\bibfnamefont {J.~J.}\ \bibnamefont {He}}, \bibinfo {author}
  {\bibfnamefont {A.~S.}\ \bibnamefont {Raja}}, \bibinfo {author}
  {\bibfnamefont {T.~Y.}\ \bibnamefont {Liu}}, \bibinfo {author} {\bibfnamefont
  {N.~J.}\ \bibnamefont {Engelsen}},\ and\ \bibinfo {author} {\bibfnamefont
  {T.~J.}\ \bibnamefont {Kippenberg}},\ }\href
  {https://doi.org/10.1038/s41467-021-21973-z} {\bibfield  {journal} {\bibinfo
  {journal} {Nature Communications}\ }\textbf {\bibinfo {volume} {12}},\
  \bibinfo {pages} {2236} (\bibinfo {year} {2021})}\BibitemShut {NoStop}%
\bibitem [{\citenamefont {Edinger}\ \emph {et~al.}(2021)\citenamefont
  {Edinger}, \citenamefont {Takabayashi}, \citenamefont {Errando-Herranz},
  \citenamefont {Khan}, \citenamefont {Sattari}, \citenamefont {Verheyen},
  \citenamefont {Bogaerts}, \citenamefont {Quack},\ and\ \citenamefont
  {Gylfason}}]{Edinger2021}%
  \BibitemOpen
  \bibfield  {author} {\bibinfo {author} {\bibfnamefont {P.}~\bibnamefont
  {Edinger}}, \bibinfo {author} {\bibfnamefont {A.~Y.}\ \bibnamefont
  {Takabayashi}}, \bibinfo {author} {\bibfnamefont {C.}~\bibnamefont
  {Errando-Herranz}}, \bibinfo {author} {\bibfnamefont {U.}~\bibnamefont
  {Khan}}, \bibinfo {author} {\bibfnamefont {H.}~\bibnamefont {Sattari}},
  \bibinfo {author} {\bibfnamefont {P.}~\bibnamefont {Verheyen}}, \bibinfo
  {author} {\bibfnamefont {W.}~\bibnamefont {Bogaerts}}, \bibinfo {author}
  {\bibfnamefont {N.}~\bibnamefont {Quack}},\ and\ \bibinfo {author}
  {\bibfnamefont {K.~B.}\ \bibnamefont {Gylfason}},\ }\href
  {https://doi.org/10.1364/Ol.436288} {\bibfield  {journal} {\bibinfo
  {journal} {Optics Letters}\ }\textbf {\bibinfo {volume} {46}},\ \bibinfo
  {pages} {5671} (\bibinfo {year} {2021})}\BibitemShut {NoStop}%
\bibitem [{\citenamefont {Gyger}\ \emph {et~al.}(2021)\citenamefont {Gyger},
  \citenamefont {Zichi}, \citenamefont {Schweickert}, \citenamefont {Elshaari},
  \citenamefont {Steinhauer}, \citenamefont {da~Silva}, \citenamefont
  {Rastelli}, \citenamefont {Zwiller}, \citenamefont {Jons},\ and\
  \citenamefont {Errando-Herranz}}]{Gyger2021}%
  \BibitemOpen
  \bibfield  {author} {\bibinfo {author} {\bibfnamefont {S.}~\bibnamefont
  {Gyger}}, \bibinfo {author} {\bibfnamefont {J.}~\bibnamefont {Zichi}},
  \bibinfo {author} {\bibfnamefont {L.}~\bibnamefont {Schweickert}}, \bibinfo
  {author} {\bibfnamefont {A.~W.}\ \bibnamefont {Elshaari}}, \bibinfo {author}
  {\bibfnamefont {S.}~\bibnamefont {Steinhauer}}, \bibinfo {author}
  {\bibfnamefont {S.~F.~C.}\ \bibnamefont {da~Silva}}, \bibinfo {author}
  {\bibfnamefont {A.}~\bibnamefont {Rastelli}}, \bibinfo {author}
  {\bibfnamefont {V.}~\bibnamefont {Zwiller}}, \bibinfo {author} {\bibfnamefont
  {K.~D.}\ \bibnamefont {Jons}},\ and\ \bibinfo {author} {\bibfnamefont
  {C.}~\bibnamefont {Errando-Herranz}},\ }\href
  {https://doi.org/10.1038/s41467-021-21624-3} {\bibfield  {journal} {\bibinfo
  {journal} {Nature Communications}\ }\textbf {\bibinfo {volume} {12}},\
  \bibinfo {pages} {1408} (\bibinfo {year} {2021})}\BibitemShut {NoStop}%
\bibitem [{\citenamefont {Chakraborty}\ \emph {et~al.}(2020)\citenamefont
  {Chakraborty}, \citenamefont {Carolan}, \citenamefont {Clark}, \citenamefont
  {Bunandar}, \citenamefont {Gilbert}, \citenamefont {Notaros}, \citenamefont
  {Watts},\ and\ \citenamefont {Englund}}]{Chakraborty2020}%
  \BibitemOpen
  \bibfield  {author} {\bibinfo {author} {\bibfnamefont {U.}~\bibnamefont
  {Chakraborty}}, \bibinfo {author} {\bibfnamefont {J.}~\bibnamefont
  {Carolan}}, \bibinfo {author} {\bibfnamefont {G.}~\bibnamefont {Clark}},
  \bibinfo {author} {\bibfnamefont {D.}~\bibnamefont {Bunandar}}, \bibinfo
  {author} {\bibfnamefont {G.}~\bibnamefont {Gilbert}}, \bibinfo {author}
  {\bibfnamefont {J.}~\bibnamefont {Notaros}}, \bibinfo {author} {\bibfnamefont
  {M.~R.}\ \bibnamefont {Watts}},\ and\ \bibinfo {author} {\bibfnamefont
  {D.~R.}\ \bibnamefont {Englund}},\ }\href
  {https://doi.org/10.1364/Optica.403178} {\bibfield  {journal} {\bibinfo
  {journal} {Optica}\ }\textbf {\bibinfo {volume} {7}},\ \bibinfo {pages}
  {1385} (\bibinfo {year} {2020})}\BibitemShut {NoStop}%
\bibitem [{\citenamefont {Redjem}\ \emph {et~al.}(2020)\citenamefont {Redjem},
  \citenamefont {Durand}, \citenamefont {Herzig}, \citenamefont {Benali},
  \citenamefont {Pezzagna}, \citenamefont {Meijer}, \citenamefont {Kuznetsov},
  \citenamefont {Nguyen}, \citenamefont {Cueff}, \citenamefont {Gerard},
  \citenamefont {Robert-Philip}, \citenamefont {Gil}, \citenamefont {Caliste},
  \citenamefont {Pochet}, \citenamefont {Abbarchi}, \citenamefont {Jacques},
  \citenamefont {Dreau},\ and\ \citenamefont {Cassabois}}]{Redjem2020}%
  \BibitemOpen
  \bibfield  {author} {\bibinfo {author} {\bibfnamefont {W.}~\bibnamefont
  {Redjem}}, \bibinfo {author} {\bibfnamefont {A.}~\bibnamefont {Durand}},
  \bibinfo {author} {\bibfnamefont {T.}~\bibnamefont {Herzig}}, \bibinfo
  {author} {\bibfnamefont {A.}~\bibnamefont {Benali}}, \bibinfo {author}
  {\bibfnamefont {S.}~\bibnamefont {Pezzagna}}, \bibinfo {author}
  {\bibfnamefont {J.}~\bibnamefont {Meijer}}, \bibinfo {author} {\bibfnamefont
  {A.~Y.}\ \bibnamefont {Kuznetsov}}, \bibinfo {author} {\bibfnamefont {H.~S.}\
  \bibnamefont {Nguyen}}, \bibinfo {author} {\bibfnamefont {S.}~\bibnamefont
  {Cueff}}, \bibinfo {author} {\bibfnamefont {J.~M.}\ \bibnamefont {Gerard}},
  \bibinfo {author} {\bibfnamefont {I.}~\bibnamefont {Robert-Philip}}, \bibinfo
  {author} {\bibfnamefont {B.}~\bibnamefont {Gil}}, \bibinfo {author}
  {\bibfnamefont {D.}~\bibnamefont {Caliste}}, \bibinfo {author} {\bibfnamefont
  {P.}~\bibnamefont {Pochet}}, \bibinfo {author} {\bibfnamefont
  {M.}~\bibnamefont {Abbarchi}}, \bibinfo {author} {\bibfnamefont
  {V.}~\bibnamefont {Jacques}}, \bibinfo {author} {\bibfnamefont
  {A.}~\bibnamefont {Dreau}},\ and\ \bibinfo {author} {\bibfnamefont
  {G.}~\bibnamefont {Cassabois}},\ }\href
  {https://doi.org/10.1038/s41928-020-00499-0} {\bibfield  {journal} {\bibinfo
  {journal} {Nature Electronics}\ }\textbf {\bibinfo {volume} {3}},\ \bibinfo
  {pages} {738} (\bibinfo {year} {2020})}\BibitemShut {NoStop}%
\bibitem [{\citenamefont {Prabhu}\ \emph {et~al.}(2023)\citenamefont {Prabhu},
  \citenamefont {Errando-Herranz}, \citenamefont {De~Santis}, \citenamefont
  {Christen}, \citenamefont {Chen}, \citenamefont {Gerlach},\ and\
  \citenamefont {Englund}}]{Prabhu2023}%
  \BibitemOpen
  \bibfield  {author} {\bibinfo {author} {\bibfnamefont {M.}~\bibnamefont
  {Prabhu}}, \bibinfo {author} {\bibfnamefont {C.}~\bibnamefont
  {Errando-Herranz}}, \bibinfo {author} {\bibfnamefont {L.}~\bibnamefont
  {De~Santis}}, \bibinfo {author} {\bibfnamefont {I.}~\bibnamefont {Christen}},
  \bibinfo {author} {\bibfnamefont {C.}~\bibnamefont {Chen}}, \bibinfo {author}
  {\bibfnamefont {C.}~\bibnamefont {Gerlach}},\ and\ \bibinfo {author}
  {\bibfnamefont {D.}~\bibnamefont {Englund}},\ }\href
  {https://doi.org/10.1038/s41467-023-37655-x} {\bibfield  {journal} {\bibinfo
  {journal} {Nature Communications}\ }\textbf {\bibinfo {volume} {14}},\
  \bibinfo {pages} {2380} (\bibinfo {year} {2023})}\BibitemShut {NoStop}%
\bibitem [{\citenamefont {Redjem}\ \emph {et~al.}(2023)\citenamefont {Redjem},
  \citenamefont {Zhiyenbayev}, \citenamefont {Qarony}, \citenamefont {Ivanov},
  \citenamefont {Papapanos}, \citenamefont {Liu}, \citenamefont {Jhuria},
  \citenamefont {Al~Balushi}, \citenamefont {Dhuey}, \citenamefont
  {Schwartzberg}, \citenamefont {Tan}, \citenamefont {Schenkel},\ and\
  \citenamefont {Kant\'{e}}}]{Redjem2023}%
  \BibitemOpen
  \bibfield  {author} {\bibinfo {author} {\bibfnamefont {W.}~\bibnamefont
  {Redjem}}, \bibinfo {author} {\bibfnamefont {Y.}~\bibnamefont {Zhiyenbayev}},
  \bibinfo {author} {\bibfnamefont {W.}~\bibnamefont {Qarony}}, \bibinfo
  {author} {\bibfnamefont {V.}~\bibnamefont {Ivanov}}, \bibinfo {author}
  {\bibfnamefont {C.}~\bibnamefont {Papapanos}}, \bibinfo {author}
  {\bibfnamefont {W.}~\bibnamefont {Liu}}, \bibinfo {author} {\bibfnamefont
  {K.}~\bibnamefont {Jhuria}}, \bibinfo {author} {\bibfnamefont {Z.~Y.}\
  \bibnamefont {Al~Balushi}}, \bibinfo {author} {\bibfnamefont
  {S.}~\bibnamefont {Dhuey}}, \bibinfo {author} {\bibfnamefont
  {A.}~\bibnamefont {Schwartzberg}}, \bibinfo {author} {\bibfnamefont {L.~Z.}\
  \bibnamefont {Tan}}, \bibinfo {author} {\bibfnamefont {T.}~\bibnamefont
  {Schenkel}},\ and\ \bibinfo {author} {\bibfnamefont {B.}~\bibnamefont
  {Kant\'{e}}},\ }\href {https://doi.org/10.1038/s41467-023-38559-6} {\bibfield
   {journal} {\bibinfo  {journal} {Nature Communications}\ }\textbf {\bibinfo
  {volume} {14}},\ \bibinfo {pages} {3321} (\bibinfo {year}
  {2023})}\BibitemShut {NoStop}%
\bibitem [{\citenamefont {Hollenbach}\ \emph {et~al.}(2022)\citenamefont
  {Hollenbach}, \citenamefont {Klingner}, \citenamefont {Jagtap}, \citenamefont
  {Bischoff}, \citenamefont {Fowley}, \citenamefont {Kentsch}, \citenamefont
  {Hlawacek}, \citenamefont {Erbe}, \citenamefont {Abrosimov}, \citenamefont
  {Helm}, \citenamefont {Berencen},\ and\ \citenamefont
  {Astakhov}}]{Hollenbach2022}%
  \BibitemOpen
  \bibfield  {author} {\bibinfo {author} {\bibfnamefont {M.}~\bibnamefont
  {Hollenbach}}, \bibinfo {author} {\bibfnamefont {N.}~\bibnamefont
  {Klingner}}, \bibinfo {author} {\bibfnamefont {N.~S.}\ \bibnamefont
  {Jagtap}}, \bibinfo {author} {\bibfnamefont {L.}~\bibnamefont {Bischoff}},
  \bibinfo {author} {\bibfnamefont {C.}~\bibnamefont {Fowley}}, \bibinfo
  {author} {\bibfnamefont {U.}~\bibnamefont {Kentsch}}, \bibinfo {author}
  {\bibfnamefont {G.}~\bibnamefont {Hlawacek}}, \bibinfo {author}
  {\bibfnamefont {A.}~\bibnamefont {Erbe}}, \bibinfo {author} {\bibfnamefont
  {N.~V.}\ \bibnamefont {Abrosimov}}, \bibinfo {author} {\bibfnamefont
  {M.}~\bibnamefont {Helm}}, \bibinfo {author} {\bibfnamefont {Y.}~\bibnamefont
  {Berencen}},\ and\ \bibinfo {author} {\bibfnamefont {G.~V.}\ \bibnamefont
  {Astakhov}},\ }\href {https://doi.org/10.1038/s41467-022-35051-5} {\bibfield
  {journal} {\bibinfo  {journal} {Nature Communications}\ }\textbf {\bibinfo
  {volume} {13}},\ \bibinfo {pages} {7683} (\bibinfo {year}
  {2022})}\BibitemShut {NoStop}%
\bibitem [{\citenamefont {Higginbottom}\ \emph {et~al.}(2022)\citenamefont
  {Higginbottom}, \citenamefont {Kurkjian}, \citenamefont {Chartrand},
  \citenamefont {Kazemi}, \citenamefont {Brunelle}, \citenamefont {MacQuarrie},
  \citenamefont {Klein}, \citenamefont {Lee-Hone}, \citenamefont {Stacho},
  \citenamefont {Ruether}, \citenamefont {Bowness}, \citenamefont {Bergeron},
  \citenamefont {DeAbreu}, \citenamefont {Harrigan}, \citenamefont
  {Kanaganayagam}, \citenamefont {Marsden}, \citenamefont {Richards},
  \citenamefont {Stott}, \citenamefont {Roorda}, \citenamefont {Morse},
  \citenamefont {Thewalt},\ and\ \citenamefont {Simmons}}]{Higginbottom2022}%
  \BibitemOpen
  \bibfield  {author} {\bibinfo {author} {\bibfnamefont {D.~B.}\ \bibnamefont
  {Higginbottom}}, \bibinfo {author} {\bibfnamefont {A.~T.~K.}\ \bibnamefont
  {Kurkjian}}, \bibinfo {author} {\bibfnamefont {C.}~\bibnamefont {Chartrand}},
  \bibinfo {author} {\bibfnamefont {M.}~\bibnamefont {Kazemi}}, \bibinfo
  {author} {\bibfnamefont {N.~A.}\ \bibnamefont {Brunelle}}, \bibinfo {author}
  {\bibfnamefont {E.~R.}\ \bibnamefont {MacQuarrie}}, \bibinfo {author}
  {\bibfnamefont {J.~R.}\ \bibnamefont {Klein}}, \bibinfo {author}
  {\bibfnamefont {N.~R.}\ \bibnamefont {Lee-Hone}}, \bibinfo {author}
  {\bibfnamefont {J.}~\bibnamefont {Stacho}}, \bibinfo {author} {\bibfnamefont
  {M.}~\bibnamefont {Ruether}}, \bibinfo {author} {\bibfnamefont
  {C.}~\bibnamefont {Bowness}}, \bibinfo {author} {\bibfnamefont
  {L.}~\bibnamefont {Bergeron}}, \bibinfo {author} {\bibfnamefont
  {A.}~\bibnamefont {DeAbreu}}, \bibinfo {author} {\bibfnamefont {S.~R.}\
  \bibnamefont {Harrigan}}, \bibinfo {author} {\bibfnamefont {J.}~\bibnamefont
  {Kanaganayagam}}, \bibinfo {author} {\bibfnamefont {D.~W.}\ \bibnamefont
  {Marsden}}, \bibinfo {author} {\bibfnamefont {T.~S.}\ \bibnamefont
  {Richards}}, \bibinfo {author} {\bibfnamefont {L.~A.}\ \bibnamefont {Stott}},
  \bibinfo {author} {\bibfnamefont {S.}~\bibnamefont {Roorda}}, \bibinfo
  {author} {\bibfnamefont {K.~J.}\ \bibnamefont {Morse}}, \bibinfo {author}
  {\bibfnamefont {M.~L.~W.}\ \bibnamefont {Thewalt}},\ and\ \bibinfo {author}
  {\bibfnamefont {S.}~\bibnamefont {Simmons}},\ }\href
  {https://doi.org/10.1038/s41586-022-04821-y} {\bibfield  {journal} {\bibinfo
  {journal} {Nature}\ }\textbf {\bibinfo {volume} {607}},\ \bibinfo {pages}
  {266} (\bibinfo {year} {2022})}\BibitemShut {NoStop}%
\bibitem [{\citenamefont {Higginbottom}\ \emph {et~al.}(2023)\citenamefont
  {Higginbottom}, \citenamefont {Asadi}, \citenamefont {Chartrand},
  \citenamefont {Ji}, \citenamefont {Bergeron}, \citenamefont {Thewalt},
  \citenamefont {Simon},\ and\ \citenamefont {Simmons}}]{Higginbottom2023}%
  \BibitemOpen
  \bibfield  {author} {\bibinfo {author} {\bibfnamefont {D.~B.}\ \bibnamefont
  {Higginbottom}}, \bibinfo {author} {\bibfnamefont {F.~K.}\ \bibnamefont
  {Asadi}}, \bibinfo {author} {\bibfnamefont {C.}~\bibnamefont {Chartrand}},
  \bibinfo {author} {\bibfnamefont {J.~W.}\ \bibnamefont {Ji}}, \bibinfo
  {author} {\bibfnamefont {L.}~\bibnamefont {Bergeron}}, \bibinfo {author}
  {\bibfnamefont {M.~L.~W.}\ \bibnamefont {Thewalt}}, \bibinfo {author}
  {\bibfnamefont {C.}~\bibnamefont {Simon}},\ and\ \bibinfo {author}
  {\bibfnamefont {S.}~\bibnamefont {Simmons}},\ }\href
  {https://doi.org/10.1103/PRXQuantum.4.020308} {\bibfield  {journal} {\bibinfo
   {journal} {PRX Quantum}\ }\textbf {\bibinfo {volume} {4}},\ \bibinfo {pages}
  {020308} (\bibinfo {year} {2023})}\BibitemShut {NoStop}%
\bibitem [{\citenamefont {Duan}\ and\ \citenamefont {Kimble}(2004)}]{Duan2004}%
  \BibitemOpen
  \bibfield  {author} {\bibinfo {author} {\bibfnamefont {L.~M.}\ \bibnamefont
  {Duan}}\ and\ \bibinfo {author} {\bibfnamefont {H.~J.}\ \bibnamefont
  {Kimble}},\ }\href {https://doi.org/10.1103/PhysRevLett.92.127902} {\bibfield
   {journal} {\bibinfo  {journal} {Physical Review Letters}\ }\textbf {\bibinfo
  {volume} {92}},\ \bibinfo {pages} {127902} (\bibinfo {year}
  {2004})}\BibitemShut {NoStop}%
\bibitem [{\citenamefont {McAuslan}\ \emph {et~al.}(2009)\citenamefont
  {McAuslan}, \citenamefont {Longdell},\ and\ \citenamefont
  {Sellars}}]{McAuslan2009}%
  \BibitemOpen
  \bibfield  {author} {\bibinfo {author} {\bibfnamefont {D.~L.}\ \bibnamefont
  {McAuslan}}, \bibinfo {author} {\bibfnamefont {J.~J.}\ \bibnamefont
  {Longdell}},\ and\ \bibinfo {author} {\bibfnamefont {M.~J.}\ \bibnamefont
  {Sellars}},\ }\href {https://doi.org/10.1103/PhysRevA.80.062307} {\bibfield
  {journal} {\bibinfo  {journal} {Physical Review A}\ }\textbf {\bibinfo
  {volume} {80}},\ \bibinfo {pages} {062307} (\bibinfo {year}
  {2009})}\BibitemShut {NoStop}%
\bibitem [{\citenamefont {B\"ottger}\ \emph {et~al.}(2009)\citenamefont
  {B\"ottger}, \citenamefont {Thiel}, \citenamefont {Cone},\ and\ \citenamefont
  {Sun}}]{Bottger2009}%
  \BibitemOpen
  \bibfield  {author} {\bibinfo {author} {\bibfnamefont {T.}~\bibnamefont
  {B\"ottger}}, \bibinfo {author} {\bibfnamefont {C.~W.}\ \bibnamefont
  {Thiel}}, \bibinfo {author} {\bibfnamefont {R.~L.}\ \bibnamefont {Cone}},\
  and\ \bibinfo {author} {\bibfnamefont {Y.}~\bibnamefont {Sun}},\ }\href
  {https://doi.org/10.1103/PhysRevB.79.115104} {\bibfield  {journal} {\bibinfo
  {journal} {Physical Review B}\ }\textbf {\bibinfo {volume} {79}},\ \bibinfo
  {pages} {115104} (\bibinfo {year} {2009})}\BibitemShut {NoStop}%
\bibitem [{\citenamefont {{Ran{\v{c}}i{\'c}}}\ \emph
  {et~al.}(2018)\citenamefont {{Ran{\v{c}}i{\'c}}}, \citenamefont {Hedges},
  \citenamefont {Ahlefeldt},\ and\ \citenamefont {Sellars}}]{Rancic2018}%
  \BibitemOpen
  \bibfield  {author} {\bibinfo {author} {\bibfnamefont {M.}~\bibnamefont
  {{Ran{\v{c}}i{\'c}}}}, \bibinfo {author} {\bibfnamefont {M.~P.}\ \bibnamefont
  {Hedges}}, \bibinfo {author} {\bibfnamefont {R.~L.}\ \bibnamefont
  {Ahlefeldt}},\ and\ \bibinfo {author} {\bibfnamefont {M.~J.}\ \bibnamefont
  {Sellars}},\ }\href {https://doi.org/10.1038/nphys4254} {\bibfield  {journal}
  {\bibinfo  {journal} {Nature Physics}\ }\textbf {\bibinfo {volume} {14}},\
  \bibinfo {pages} {50} (\bibinfo {year} {2018})}\BibitemShut {NoStop}%
\bibitem [{\citenamefont {Le~Dantec}\ \emph {et~al.}(2021)\citenamefont
  {Le~Dantec}, \citenamefont {Rancic}, \citenamefont {Lin}, \citenamefont
  {Billaud}, \citenamefont {Ranjan}, \citenamefont {Flanigan}, \citenamefont
  {Bertaina}, \citenamefont {Chaneliere}, \citenamefont {Goldner},
  \citenamefont {Erb}, \citenamefont {Liu}, \citenamefont {Esteve},
  \citenamefont {Vion}, \citenamefont {Flurin},\ and\ \citenamefont
  {Bertet}}]{Dantec2021}%
  \BibitemOpen
  \bibfield  {author} {\bibinfo {author} {\bibfnamefont {M.}~\bibnamefont
  {Le~Dantec}}, \bibinfo {author} {\bibfnamefont {M.}~\bibnamefont {Rancic}},
  \bibinfo {author} {\bibfnamefont {S.}~\bibnamefont {Lin}}, \bibinfo {author}
  {\bibfnamefont {E.}~\bibnamefont {Billaud}}, \bibinfo {author} {\bibfnamefont
  {V.}~\bibnamefont {Ranjan}}, \bibinfo {author} {\bibfnamefont
  {D.}~\bibnamefont {Flanigan}}, \bibinfo {author} {\bibfnamefont
  {S.}~\bibnamefont {Bertaina}}, \bibinfo {author} {\bibfnamefont
  {T.}~\bibnamefont {Chaneliere}}, \bibinfo {author} {\bibfnamefont
  {P.}~\bibnamefont {Goldner}}, \bibinfo {author} {\bibfnamefont
  {A.}~\bibnamefont {Erb}}, \bibinfo {author} {\bibfnamefont {R.~B.}\
  \bibnamefont {Liu}}, \bibinfo {author} {\bibfnamefont {D.}~\bibnamefont
  {Esteve}}, \bibinfo {author} {\bibfnamefont {D.}~\bibnamefont {Vion}},
  \bibinfo {author} {\bibfnamefont {E.}~\bibnamefont {Flurin}},\ and\ \bibinfo
  {author} {\bibfnamefont {P.}~\bibnamefont {Bertet}},\ }\href
  {https://doi.org/10.1126/sciadv.abj9786} {\bibfield  {journal} {\bibinfo
  {journal} {Science Advances}\ }\textbf {\bibinfo {volume} {7}},\ \bibinfo
  {pages} {eabj9786} (\bibinfo {year} {2021})}\BibitemShut {NoStop}%
\bibitem [{\citenamefont {Williamson}\ \emph {et~al.}(2014)\citenamefont
  {Williamson}, \citenamefont {Chen},\ and\ \citenamefont
  {Longdell}}]{Williamson2014}%
  \BibitemOpen
  \bibfield  {author} {\bibinfo {author} {\bibfnamefont {L.~A.}\ \bibnamefont
  {Williamson}}, \bibinfo {author} {\bibfnamefont {Y.~H.}\ \bibnamefont
  {Chen}},\ and\ \bibinfo {author} {\bibfnamefont {J.~J.}\ \bibnamefont
  {Longdell}},\ }\href {https://doi.org/10.1103/PhysRevLett.113.203601}
  {\bibfield  {journal} {\bibinfo  {journal} {Physical Review Letters}\
  }\textbf {\bibinfo {volume} {113}},\ \bibinfo {pages} {203601} (\bibinfo
  {year} {2014})}\BibitemShut {NoStop}%
\bibitem [{\citenamefont {Fernandez-Gonzalvo}\ \emph
  {et~al.}(2015)\citenamefont {Fernandez-Gonzalvo}, \citenamefont {Chen},
  \citenamefont {Yin}, \citenamefont {Rogge},\ and\ \citenamefont
  {Longdell}}]{Fernandez2015}%
  \BibitemOpen
  \bibfield  {author} {\bibinfo {author} {\bibfnamefont {X.}~\bibnamefont
  {Fernandez-Gonzalvo}}, \bibinfo {author} {\bibfnamefont {Y.~H.}\ \bibnamefont
  {Chen}}, \bibinfo {author} {\bibfnamefont {C.~M.}\ \bibnamefont {Yin}},
  \bibinfo {author} {\bibfnamefont {S.}~\bibnamefont {Rogge}},\ and\ \bibinfo
  {author} {\bibfnamefont {J.~J.}\ \bibnamefont {Longdell}},\ }\href
  {https://doi.org/10.1103/PhysRevA.92.062313} {\bibfield  {journal} {\bibinfo
  {journal} {Physical Review A}\ }\textbf {\bibinfo {volume} {92}},\ \bibinfo
  {pages} {062313} (\bibinfo {year} {2015})}\BibitemShut {NoStop}%
\bibitem [{\citenamefont {Fernandez-Gonzalvo}\ \emph
  {et~al.}(2019)\citenamefont {Fernandez-Gonzalvo}, \citenamefont {Horvath},
  \citenamefont {Chen},\ and\ \citenamefont {Longdell}}]{Fernandez2019}%
  \BibitemOpen
  \bibfield  {author} {\bibinfo {author} {\bibfnamefont {X.}~\bibnamefont
  {Fernandez-Gonzalvo}}, \bibinfo {author} {\bibfnamefont {S.~P.}\ \bibnamefont
  {Horvath}}, \bibinfo {author} {\bibfnamefont {Y.~H.}\ \bibnamefont {Chen}},\
  and\ \bibinfo {author} {\bibfnamefont {J.~J.}\ \bibnamefont {Longdell}},\
  }\href {https://doi.org/10.1103/PhysRevA.100.033807} {\bibfield  {journal}
  {\bibinfo  {journal} {Physical Review A}\ }\textbf {\bibinfo {volume}
  {100}},\ \bibinfo {pages} {033807} (\bibinfo {year} {2019})}\BibitemShut
  {NoStop}%
\bibitem [{\citenamefont {Rochman}\ \emph {et~al.}(2023)\citenamefont
  {Rochman}, \citenamefont {Xie}, \citenamefont {Bartholomew}, \citenamefont
  {Schwab},\ and\ \citenamefont {Faraon}}]{Rochman2023}%
  \BibitemOpen
  \bibfield  {author} {\bibinfo {author} {\bibfnamefont {J.}~\bibnamefont
  {Rochman}}, \bibinfo {author} {\bibfnamefont {T.}~\bibnamefont {Xie}},
  \bibinfo {author} {\bibfnamefont {J.~G.}\ \bibnamefont {Bartholomew}},
  \bibinfo {author} {\bibfnamefont {K.~C.}\ \bibnamefont {Schwab}},\ and\
  \bibinfo {author} {\bibfnamefont {A.}~\bibnamefont {Faraon}},\ }\href
  {https://doi.org/10.1038/s41467-023-36799-0} {\bibfield  {journal} {\bibinfo
  {journal} {Nature Communications}\ }\textbf {\bibinfo {volume} {14}},\
  \bibinfo {pages} {1153} (\bibinfo {year} {2023})}\BibitemShut {NoStop}%
\bibitem [{\citenamefont {{Ourari}}\ \emph {et~al.}(2023)\citenamefont
  {{Ourari}}, \citenamefont {{Dusanowski}}, \citenamefont {{Horvath}},
  \citenamefont {{Uysal}}, \citenamefont {{Phenicie}}, \citenamefont
  {{Stevenson}}, \citenamefont {{Raha}}, \citenamefont {{Chen}}, \citenamefont
  {{Cava}}, \citenamefont {{de Leon}},\ and\ \citenamefont
  {{Thompson}}}]{Ourari2023}%
  \BibitemOpen
  \bibfield  {author} {\bibinfo {author} {\bibfnamefont {S.}~\bibnamefont
  {{Ourari}}}, \bibinfo {author} {\bibfnamefont {{\L}.}~\bibnamefont
  {{Dusanowski}}}, \bibinfo {author} {\bibfnamefont {S.~P.}\ \bibnamefont
  {{Horvath}}}, \bibinfo {author} {\bibfnamefont {M.~T.}\ \bibnamefont
  {{Uysal}}}, \bibinfo {author} {\bibfnamefont {C.~M.}\ \bibnamefont
  {{Phenicie}}}, \bibinfo {author} {\bibfnamefont {P.}~\bibnamefont
  {{Stevenson}}}, \bibinfo {author} {\bibfnamefont {M.}~\bibnamefont {{Raha}}},
  \bibinfo {author} {\bibfnamefont {S.}~\bibnamefont {{Chen}}}, \bibinfo
  {author} {\bibfnamefont {R.~J.}\ \bibnamefont {{Cava}}}, \bibinfo {author}
  {\bibfnamefont {N.~P.}\ \bibnamefont {{de Leon}}},\ and\ \bibinfo {author}
  {\bibfnamefont {J.~D.}\ \bibnamefont {{Thompson}}},\ }\href
  {https://doi.org/10.48550/arXiv.2301.03564} {\bibfield  {journal} {\bibinfo
  {journal} {arXiv e-prints}\ ,\ \bibinfo {eid} {arXiv:2301.03564}} (\bibinfo
  {year} {2023})},\ \Eprint {https://arxiv.org/abs/2301.03564}
  {arXiv:2301.03564 [quant-ph]} \BibitemShut {NoStop}%
\bibitem [{\citenamefont {Raha}\ \emph {et~al.}(2020)\citenamefont {Raha},
  \citenamefont {Chen}, \citenamefont {Phenicie}, \citenamefont {Ourari},
  \citenamefont {Dibos},\ and\ \citenamefont {Thompson}}]{Raha2020}%
  \BibitemOpen
  \bibfield  {author} {\bibinfo {author} {\bibfnamefont {M.}~\bibnamefont
  {Raha}}, \bibinfo {author} {\bibfnamefont {S.~T.}\ \bibnamefont {Chen}},
  \bibinfo {author} {\bibfnamefont {C.~M.}\ \bibnamefont {Phenicie}}, \bibinfo
  {author} {\bibfnamefont {S.}~\bibnamefont {Ourari}}, \bibinfo {author}
  {\bibfnamefont {A.~M.}\ \bibnamefont {Dibos}},\ and\ \bibinfo {author}
  {\bibfnamefont {J.~D.}\ \bibnamefont {Thompson}},\ }\href
  {https://doi.org/10.1038/s41467-020-15138-7} {\bibfield  {journal} {\bibinfo
  {journal} {Nature Communications}\ }\textbf {\bibinfo {volume} {11}},\
  \bibinfo {pages} {1605} (\bibinfo {year} {2020})}\BibitemShut {NoStop}%
\bibitem [{\citenamefont {Yang}\ \emph {et~al.}(2023)\citenamefont {Yang},
  \citenamefont {Wang}, \citenamefont {Shen}, \citenamefont {Xie},\ and\
  \citenamefont {Tang}}]{Yang2023}%
  \BibitemOpen
  \bibfield  {author} {\bibinfo {author} {\bibfnamefont {L.~K.}\ \bibnamefont
  {Yang}}, \bibinfo {author} {\bibfnamefont {S.~H.}\ \bibnamefont {Wang}},
  \bibinfo {author} {\bibfnamefont {M.~H.}\ \bibnamefont {Shen}}, \bibinfo
  {author} {\bibfnamefont {J.~C.}\ \bibnamefont {Xie}},\ and\ \bibinfo {author}
  {\bibfnamefont {H.~X.}\ \bibnamefont {Tang}},\ }\href
  {https://doi.org/10.1038/s41467-023-37513-w} {\bibfield  {journal} {\bibinfo
  {journal} {Nature Communications}\ }\textbf {\bibinfo {volume} {14}},\
  \bibinfo {pages} {1718} (\bibinfo {year} {2023})}\BibitemShut {NoStop}%
\bibitem [{\citenamefont {{Wang}}\ \emph {et~al.}(2023)\citenamefont {{Wang}},
  \citenamefont {{Balembois}}, \citenamefont {{Ran{\v{c}}i{\'c}}},
  \citenamefont {{Billaud}}, \citenamefont {{Le Dantec}}, \citenamefont
  {{Ferrier}}, \citenamefont {{Goldner}}, \citenamefont {{Bertaina}},
  \citenamefont {{Chaneli{\`e}re}}, \citenamefont {{Est{\`e}ve}}, \citenamefont
  {{Vion}}, \citenamefont {{Bertet}},\ and\ \citenamefont
  {{Flurin}}}]{Wang2023}%
  \BibitemOpen
  \bibfield  {author} {\bibinfo {author} {\bibfnamefont {Z.}~\bibnamefont
  {{Wang}}}, \bibinfo {author} {\bibfnamefont {L.}~\bibnamefont {{Balembois}}},
  \bibinfo {author} {\bibfnamefont {M.}~\bibnamefont {{Ran{\v{c}}i{\'c}}}},
  \bibinfo {author} {\bibfnamefont {E.}~\bibnamefont {{Billaud}}}, \bibinfo
  {author} {\bibfnamefont {M.}~\bibnamefont {{Le Dantec}}}, \bibinfo {author}
  {\bibfnamefont {A.}~\bibnamefont {{Ferrier}}}, \bibinfo {author}
  {\bibfnamefont {P.}~\bibnamefont {{Goldner}}}, \bibinfo {author}
  {\bibfnamefont {S.}~\bibnamefont {{Bertaina}}}, \bibinfo {author}
  {\bibfnamefont {T.}~\bibnamefont {{Chaneli{\`e}re}}}, \bibinfo {author}
  {\bibfnamefont {D.}~\bibnamefont {{Est{\`e}ve}}}, \bibinfo {author}
  {\bibfnamefont {D.}~\bibnamefont {{Vion}}}, \bibinfo {author} {\bibfnamefont
  {P.}~\bibnamefont {{Bertet}}},\ and\ \bibinfo {author} {\bibfnamefont
  {E.}~\bibnamefont {{Flurin}}},\ }\href
  {https://doi.org/10.1038/s41586-023-06097-2} {\bibfield  {journal} {\bibinfo
  {journal} {Nature}\ }\textbf {\bibinfo {volume} {619}},\ \bibinfo {pages}
  {276} (\bibinfo {year} {2023})}\BibitemShut {NoStop}%
\bibitem [{\citenamefont {Berkman}\ \emph {et~al.}(2023)\citenamefont
  {Berkman}, \citenamefont {Lyasota}, \citenamefont {de~Boo}, \citenamefont
  {Bartholomew}, \citenamefont {Johnson}, \citenamefont {McCallum},
  \citenamefont {Xu}, \citenamefont {Xie}, \citenamefont {Ahlefeldt},
  \citenamefont {Sellars}, \citenamefont {Yin},\ and\ \citenamefont
  {Rogge}}]{Berkman2023}%
  \BibitemOpen
  \bibfield  {author} {\bibinfo {author} {\bibfnamefont {I.~R.}\ \bibnamefont
  {Berkman}}, \bibinfo {author} {\bibfnamefont {A.}~\bibnamefont {Lyasota}},
  \bibinfo {author} {\bibfnamefont {G.~G.}\ \bibnamefont {de~Boo}}, \bibinfo
  {author} {\bibfnamefont {J.~G.}\ \bibnamefont {Bartholomew}}, \bibinfo
  {author} {\bibfnamefont {B.~C.}\ \bibnamefont {Johnson}}, \bibinfo {author}
  {\bibfnamefont {J.~C.}\ \bibnamefont {McCallum}}, \bibinfo {author}
  {\bibfnamefont {B.-B.}\ \bibnamefont {Xu}}, \bibinfo {author} {\bibfnamefont
  {S.}~\bibnamefont {Xie}}, \bibinfo {author} {\bibfnamefont {R.~L.}\
  \bibnamefont {Ahlefeldt}}, \bibinfo {author} {\bibfnamefont {M.~J.}\
  \bibnamefont {Sellars}}, \bibinfo {author} {\bibfnamefont {C.}~\bibnamefont
  {Yin}},\ and\ \bibinfo {author} {\bibfnamefont {S.}~\bibnamefont {Rogge}},\
  }\href {https://doi.org/10.1103/PhysRevApplied.19.014037} {\bibfield
  {journal} {\bibinfo  {journal} {Phys. Rev. Appl.}\ }\textbf {\bibinfo
  {volume} {19}},\ \bibinfo {pages} {014037} (\bibinfo {year}
  {2023})}\BibitemShut {NoStop}%
\bibitem [{\citenamefont {Gritsch}\ \emph {et~al.}(2022)\citenamefont
  {Gritsch}, \citenamefont {Weiss}, \citenamefont {Fruh}, \citenamefont
  {Rinner},\ and\ \citenamefont {Reiserer}}]{Gritsch2022}%
  \BibitemOpen
  \bibfield  {author} {\bibinfo {author} {\bibfnamefont {A.}~\bibnamefont
  {Gritsch}}, \bibinfo {author} {\bibfnamefont {L.}~\bibnamefont {Weiss}},
  \bibinfo {author} {\bibfnamefont {J.}~\bibnamefont {Fruh}}, \bibinfo {author}
  {\bibfnamefont {S.}~\bibnamefont {Rinner}},\ and\ \bibinfo {author}
  {\bibfnamefont {A.}~\bibnamefont {Reiserer}},\ }\href
  {https://doi.org/10.1103/PhysRevX.12.041009} {\bibfield  {journal} {\bibinfo
  {journal} {Physical Review X}\ }\textbf {\bibinfo {volume} {12}},\ \bibinfo
  {pages} {041009} (\bibinfo {year} {2022})}\BibitemShut {NoStop}%
\bibitem [{\citenamefont {Hughes}\ \emph {et~al.}(2021)\citenamefont {Hughes},
  \citenamefont {Panjwani}, \citenamefont {Urdampilleta}, \citenamefont
  {Homewood}, \citenamefont {Murdin},\ and\ \citenamefont
  {Carey}}]{Hughes2021}%
  \BibitemOpen
  \bibfield  {author} {\bibinfo {author} {\bibfnamefont {M.~A.}\ \bibnamefont
  {Hughes}}, \bibinfo {author} {\bibfnamefont {N.~A.}\ \bibnamefont
  {Panjwani}}, \bibinfo {author} {\bibfnamefont {M.}~\bibnamefont
  {Urdampilleta}}, \bibinfo {author} {\bibfnamefont {K.~P.}\ \bibnamefont
  {Homewood}}, \bibinfo {author} {\bibfnamefont {B.}~\bibnamefont {Murdin}},\
  and\ \bibinfo {author} {\bibfnamefont {J.~D.}\ \bibnamefont {Carey}},\ }\href
  {https://doi.org/10.1063/5.0046904} {\bibfield  {journal} {\bibinfo
  {journal} {Applied Physics Letters}\ }\textbf {\bibinfo {volume} {118}},\
  \bibinfo {pages} {194001} (\bibinfo {year} {2021})}\BibitemShut {NoStop}%
\bibitem [{\citenamefont {Kenyon}(2005)}]{Kenyon2005}%
  \BibitemOpen
  \bibfield  {author} {\bibinfo {author} {\bibfnamefont {A.~J.}\ \bibnamefont
  {Kenyon}},\ }\href {https://doi.org/10.1088/0268-1242/20/12/R02} {\bibfield
  {journal} {\bibinfo  {journal} {Semiconductor Science and Technology}\
  }\textbf {\bibinfo {volume} {20}},\ \bibinfo {pages} {R65} (\bibinfo {year}
  {2005})}\BibitemShut {NoStop}%
\bibitem [{\citenamefont {Przybyli\ifmmode~\acute{n}\else \'{n}\fi{}ska}\ \emph
  {et~al.}(1996)\citenamefont {Przybyli\ifmmode~\acute{n}\else \'{n}\fi{}ska},
  \citenamefont {Jantsch}, \citenamefont {Suprun-Belevitch}, \citenamefont
  {Stepikhova}, \citenamefont {Palmetshofer}, \citenamefont {Hendorfer},
  \citenamefont {Kozanecki}, \citenamefont {Wilson},\ and\ \citenamefont
  {Sealy}}]{Przybylinska1996}%
  \BibitemOpen
  \bibfield  {author} {\bibinfo {author} {\bibfnamefont {H.}~\bibnamefont
  {Przybyli\ifmmode~\acute{n}\else \'{n}\fi{}ska}}, \bibinfo {author}
  {\bibfnamefont {W.}~\bibnamefont {Jantsch}}, \bibinfo {author} {\bibfnamefont
  {Y.}~\bibnamefont {Suprun-Belevitch}}, \bibinfo {author} {\bibfnamefont
  {M.}~\bibnamefont {Stepikhova}}, \bibinfo {author} {\bibfnamefont
  {L.}~\bibnamefont {Palmetshofer}}, \bibinfo {author} {\bibfnamefont
  {G.}~\bibnamefont {Hendorfer}}, \bibinfo {author} {\bibfnamefont
  {A.}~\bibnamefont {Kozanecki}}, \bibinfo {author} {\bibfnamefont {R.~J.}\
  \bibnamefont {Wilson}},\ and\ \bibinfo {author} {\bibfnamefont {B.~J.}\
  \bibnamefont {Sealy}},\ }\href {https://doi.org/10.1103/PhysRevB.54.2532}
  {\bibfield  {journal} {\bibinfo  {journal} {Physical Review B}\ }\textbf
  {\bibinfo {volume} {54}},\ \bibinfo {pages} {2532} (\bibinfo {year}
  {1996})}\BibitemShut {NoStop}%
\bibitem [{\citenamefont {Weiss}\ \emph {et~al.}(2021)\citenamefont {Weiss},
  \citenamefont {Gritsch}, \citenamefont {Merkel},\ and\ \citenamefont
  {Reiserer}}]{Weiss2021}%
  \BibitemOpen
  \bibfield  {author} {\bibinfo {author} {\bibfnamefont {L.}~\bibnamefont
  {Weiss}}, \bibinfo {author} {\bibfnamefont {A.}~\bibnamefont {Gritsch}},
  \bibinfo {author} {\bibfnamefont {B.}~\bibnamefont {Merkel}},\ and\ \bibinfo
  {author} {\bibfnamefont {A.}~\bibnamefont {Reiserer}},\ }\href
  {https://doi.org/10.1364/Optica.413330} {\bibfield  {journal} {\bibinfo
  {journal} {Optica}\ }\textbf {\bibinfo {volume} {8}},\ \bibinfo {pages} {40}
  (\bibinfo {year} {2021})}\BibitemShut {NoStop}%
\bibitem [{\citenamefont {Macfarlane}\ \emph {et~al.}(1992)\citenamefont
  {Macfarlane}, \citenamefont {Cassanho},\ and\ \citenamefont
  {Meltzer}}]{Macfarlane1992}%
  \BibitemOpen
  \bibfield  {author} {\bibinfo {author} {\bibfnamefont {R.~M.}\ \bibnamefont
  {Macfarlane}}, \bibinfo {author} {\bibfnamefont {A.}~\bibnamefont
  {Cassanho}},\ and\ \bibinfo {author} {\bibfnamefont {R.~S.}\ \bibnamefont
  {Meltzer}},\ }\href {https://doi.org/10.1103/PhysRevLett.69.542} {\bibfield
  {journal} {\bibinfo  {journal} {Physical Review Letters}\ }\textbf {\bibinfo
  {volume} {69}},\ \bibinfo {pages} {542} (\bibinfo {year} {1992})}\BibitemShut
  {NoStop}%
\bibitem [{\citenamefont {Bjorklund}(1988)}]{moerner1988}%
  \BibitemOpen
  \bibfield  {author} {\bibinfo {author} {\bibfnamefont {G.~C.}\ \bibnamefont
  {Bjorklund}},\ }in\ \href {https://doi.org/10.1007/978-3-642-83290-1} {\emph
  {\bibinfo {booktitle} {Persistent Spectral Hole-Burning: Science and
  Applications}}},\ Vol.~\bibinfo {volume} {1},\ \bibinfo {editor} {edited by\
  \bibinfo {editor} {\bibfnamefont {W.~E.}\ \bibnamefont {Moerner}}}\ (\bibinfo
   {publisher} {Springer-Verlag Berlin Heidelberg},\ \bibinfo {year} {1988})\
  p.~\bibinfo {pages} {5}\BibitemShut {NoStop}%
\bibitem [{\citenamefont {Vahapoglu}\ \emph {et~al.}(2022)\citenamefont
  {Vahapoglu}, \citenamefont {Slack-Smith}, \citenamefont {Leon}, \citenamefont
  {Lim}, \citenamefont {Hudson}, \citenamefont {Day}, \citenamefont
  {Cifuentes}, \citenamefont {Tanttu}, \citenamefont {Yang}, \citenamefont
  {Saraiva}, \citenamefont {Abrosimov}, \citenamefont {Pohl}, \citenamefont
  {Thewalt}, \citenamefont {Laucht}, \citenamefont {Dzurak},\ and\
  \citenamefont {Pla}}]{Vahapoglu2022}%
  \BibitemOpen
  \bibfield  {author} {\bibinfo {author} {\bibfnamefont {E.}~\bibnamefont
  {Vahapoglu}}, \bibinfo {author} {\bibfnamefont {J.~P.}\ \bibnamefont
  {Slack-Smith}}, \bibinfo {author} {\bibfnamefont {R.~C.~C.}\ \bibnamefont
  {Leon}}, \bibinfo {author} {\bibfnamefont {W.~H.}\ \bibnamefont {Lim}},
  \bibinfo {author} {\bibfnamefont {F.~E.}\ \bibnamefont {Hudson}}, \bibinfo
  {author} {\bibfnamefont {T.}~\bibnamefont {Day}}, \bibinfo {author}
  {\bibfnamefont {J.~D.}\ \bibnamefont {Cifuentes}}, \bibinfo {author}
  {\bibfnamefont {T.}~\bibnamefont {Tanttu}}, \bibinfo {author} {\bibfnamefont
  {C.~H.}\ \bibnamefont {Yang}}, \bibinfo {author} {\bibfnamefont
  {A.}~\bibnamefont {Saraiva}}, \bibinfo {author} {\bibfnamefont {N.~V.}\
  \bibnamefont {Abrosimov}}, \bibinfo {author} {\bibfnamefont {H.~J.}\
  \bibnamefont {Pohl}}, \bibinfo {author} {\bibfnamefont {M.~L.~W.}\
  \bibnamefont {Thewalt}}, \bibinfo {author} {\bibfnamefont {A.}~\bibnamefont
  {Laucht}}, \bibinfo {author} {\bibfnamefont {A.~S.}\ \bibnamefont {Dzurak}},\
  and\ \bibinfo {author} {\bibfnamefont {J.~J.}\ \bibnamefont {Pla}},\ }\href
  {https://doi.org/10.1038/s41534-022-00645-w} {\bibfield  {journal} {\bibinfo
  {journal} {npj Quantum Information}\ }\textbf {\bibinfo {volume} {8}},\
  \bibinfo {pages} {126} (\bibinfo {year} {2022})}\BibitemShut {NoStop}%
\bibitem [{\citenamefont {Undseth}\ \emph {et~al.}(2023)\citenamefont
  {Undseth}, \citenamefont {Xue}, \citenamefont {Mehmandoost}, \citenamefont
  {Rimbach-Russ}, \citenamefont {Eendebak}, \citenamefont {Samkharadze},
  \citenamefont {Sammak}, \citenamefont {Dobrovitski}, \citenamefont
  {Scappucci},\ and\ \citenamefont {Vandersypen}}]{Undseth2023}%
  \BibitemOpen
  \bibfield  {author} {\bibinfo {author} {\bibfnamefont {B.}~\bibnamefont
  {Undseth}}, \bibinfo {author} {\bibfnamefont {X.}~\bibnamefont {Xue}},
  \bibinfo {author} {\bibfnamefont {M.}~\bibnamefont {Mehmandoost}}, \bibinfo
  {author} {\bibfnamefont {M.}~\bibnamefont {Rimbach-Russ}}, \bibinfo {author}
  {\bibfnamefont {P.~T.}\ \bibnamefont {Eendebak}}, \bibinfo {author}
  {\bibfnamefont {N.}~\bibnamefont {Samkharadze}}, \bibinfo {author}
  {\bibfnamefont {A.}~\bibnamefont {Sammak}}, \bibinfo {author} {\bibfnamefont
  {V.~V.}\ \bibnamefont {Dobrovitski}}, \bibinfo {author} {\bibfnamefont
  {G.}~\bibnamefont {Scappucci}},\ and\ \bibinfo {author} {\bibfnamefont
  {L.~M.}\ \bibnamefont {Vandersypen}},\ }\href
  {https://doi.org/10.1103/PhysRevApplied.19.044078} {\bibfield  {journal}
  {\bibinfo  {journal} {Phys. Rev. Appl.}\ }\textbf {\bibinfo {volume} {19}},\
  \bibinfo {pages} {044078} (\bibinfo {year} {2023})}\BibitemShut {NoStop}%
\bibitem [{\citenamefont {Yoneda}\ \emph {et~al.}(2014)\citenamefont {Yoneda},
  \citenamefont {Otsuka}, \citenamefont {Nakajima}, \citenamefont {Takakura},
  \citenamefont {Obata}, \citenamefont {Pioro-Ladriere}, \citenamefont {Lu},
  \citenamefont {Palmstrom}, \citenamefont {Gossard},\ and\ \citenamefont
  {Tarucha}}]{Yoneda2014}%
  \BibitemOpen
  \bibfield  {author} {\bibinfo {author} {\bibfnamefont {J.}~\bibnamefont
  {Yoneda}}, \bibinfo {author} {\bibfnamefont {T.}~\bibnamefont {Otsuka}},
  \bibinfo {author} {\bibfnamefont {T.}~\bibnamefont {Nakajima}}, \bibinfo
  {author} {\bibfnamefont {T.}~\bibnamefont {Takakura}}, \bibinfo {author}
  {\bibfnamefont {T.}~\bibnamefont {Obata}}, \bibinfo {author} {\bibfnamefont
  {M.}~\bibnamefont {Pioro-Ladriere}}, \bibinfo {author} {\bibfnamefont
  {H.}~\bibnamefont {Lu}}, \bibinfo {author} {\bibfnamefont {C.~J.}\
  \bibnamefont {Palmstrom}}, \bibinfo {author} {\bibfnamefont {A.~C.}\
  \bibnamefont {Gossard}},\ and\ \bibinfo {author} {\bibfnamefont
  {S.}~\bibnamefont {Tarucha}},\ }\href
  {https://doi.org/10.1103/PhysRevLett.113.267601} {\bibfield  {journal}
  {\bibinfo  {journal} {Physical Review Letters}\ }\textbf {\bibinfo {volume}
  {113}},\ \bibinfo {pages} {267601} (\bibinfo {year} {2014})}\BibitemShut
  {NoStop}%
\bibitem [{\citenamefont {Takeda}\ \emph {et~al.}(2016)\citenamefont {Takeda},
  \citenamefont {Kamioka}, \citenamefont {Otsuka}, \citenamefont {Yoneda},
  \citenamefont {Nakajima}, \citenamefont {Delbecq}, \citenamefont {Amaha},
  \citenamefont {Allison}, \citenamefont {Kodera}, \citenamefont {Oda},\ and\
  \citenamefont {Tarucha}}]{Takeda2016}%
  \BibitemOpen
  \bibfield  {author} {\bibinfo {author} {\bibfnamefont {K.}~\bibnamefont
  {Takeda}}, \bibinfo {author} {\bibfnamefont {J.}~\bibnamefont {Kamioka}},
  \bibinfo {author} {\bibfnamefont {T.}~\bibnamefont {Otsuka}}, \bibinfo
  {author} {\bibfnamefont {J.}~\bibnamefont {Yoneda}}, \bibinfo {author}
  {\bibfnamefont {T.}~\bibnamefont {Nakajima}}, \bibinfo {author}
  {\bibfnamefont {M.~R.}\ \bibnamefont {Delbecq}}, \bibinfo {author}
  {\bibfnamefont {S.}~\bibnamefont {Amaha}}, \bibinfo {author} {\bibfnamefont
  {G.}~\bibnamefont {Allison}}, \bibinfo {author} {\bibfnamefont
  {T.}~\bibnamefont {Kodera}}, \bibinfo {author} {\bibfnamefont
  {S.}~\bibnamefont {Oda}},\ and\ \bibinfo {author} {\bibfnamefont
  {S.}~\bibnamefont {Tarucha}},\ }\href
  {https://doi.org/10.1126/sciadv.1600694} {\bibfield  {journal} {\bibinfo
  {journal} {Science Advances}\ }\textbf {\bibinfo {volume} {2}},\ \bibinfo
  {pages} {e1600694} (\bibinfo {year} {2016})}\BibitemShut {NoStop}%
\bibitem [{\citenamefont {Nakajima}\ \emph {et~al.}(2020)\citenamefont
  {Nakajima}, \citenamefont {Noiri}, \citenamefont {Kawasaki}, \citenamefont
  {Yoneda}, \citenamefont {Stano}, \citenamefont {Amaha}, \citenamefont
  {Otsuka}, \citenamefont {Takeda}, \citenamefont {Delbecq}, \citenamefont
  {Allison}, \citenamefont {Ludwig}, \citenamefont {Wieck}, \citenamefont
  {Loss},\ and\ \citenamefont {Tarucha}}]{Nakajima2020}%
  \BibitemOpen
  \bibfield  {author} {\bibinfo {author} {\bibfnamefont {T.}~\bibnamefont
  {Nakajima}}, \bibinfo {author} {\bibfnamefont {A.}~\bibnamefont {Noiri}},
  \bibinfo {author} {\bibfnamefont {K.}~\bibnamefont {Kawasaki}}, \bibinfo
  {author} {\bibfnamefont {J.}~\bibnamefont {Yoneda}}, \bibinfo {author}
  {\bibfnamefont {P.}~\bibnamefont {Stano}}, \bibinfo {author} {\bibfnamefont
  {S.}~\bibnamefont {Amaha}}, \bibinfo {author} {\bibfnamefont
  {T.}~\bibnamefont {Otsuka}}, \bibinfo {author} {\bibfnamefont
  {K.}~\bibnamefont {Takeda}}, \bibinfo {author} {\bibfnamefont {M.~R.}\
  \bibnamefont {Delbecq}}, \bibinfo {author} {\bibfnamefont {G.}~\bibnamefont
  {Allison}}, \bibinfo {author} {\bibfnamefont {A.}~\bibnamefont {Ludwig}},
  \bibinfo {author} {\bibfnamefont {A.~D.}\ \bibnamefont {Wieck}}, \bibinfo
  {author} {\bibfnamefont {D.}~\bibnamefont {Loss}},\ and\ \bibinfo {author}
  {\bibfnamefont {S.}~\bibnamefont {Tarucha}},\ }\href
  {https://doi.org/10.1103/PhysRevX.10.011060} {\bibfield  {journal} {\bibinfo
  {journal} {Physical Review X}\ }\textbf {\bibinfo {volume} {10}},\ \bibinfo
  {pages} {011060} (\bibinfo {year} {2020})}\BibitemShut {NoStop}%
\bibitem [{\citenamefont {Hahn}(1950)}]{Hahn1950}%
  \BibitemOpen
  \bibfield  {author} {\bibinfo {author} {\bibfnamefont {E.~L.}\ \bibnamefont
  {Hahn}},\ }\href {https://doi.org/10.1103/PhysRev.80.580} {\bibfield
  {journal} {\bibinfo  {journal} {Physical Review}\ }\textbf {\bibinfo {volume}
  {80}},\ \bibinfo {pages} {580} (\bibinfo {year} {1950})}\BibitemShut
  {NoStop}%
\bibitem [{\citenamefont {Chiba}\ and\ \citenamefont
  {Hirai}(1972)}]{Chiba1972}%
  \BibitemOpen
  \bibfield  {author} {\bibinfo {author} {\bibfnamefont {M.}~\bibnamefont
  {Chiba}}\ and\ \bibinfo {author} {\bibfnamefont {A.}~\bibnamefont {Hirai}},\
  }\href {https://doi.org/Doi 10.1143/Jpsj.33.730} {\bibfield  {journal}
  {\bibinfo  {journal} {Journal of the Physical Society of Japan}\ }\textbf
  {\bibinfo {volume} {33}},\ \bibinfo {pages} {730} (\bibinfo {year}
  {1972})}\BibitemShut {NoStop}%
\bibitem [{\citenamefont {Tyryshkin}\ \emph {et~al.}(2006)\citenamefont
  {Tyryshkin}, \citenamefont {Morton}, \citenamefont {Benjamin}, \citenamefont
  {Ardavan}, \citenamefont {Briggs}, \citenamefont {Ager},\ and\ \citenamefont
  {Lyon}}]{Tyryshkin2006}%
  \BibitemOpen
  \bibfield  {author} {\bibinfo {author} {\bibfnamefont {A.~M.}\ \bibnamefont
  {Tyryshkin}}, \bibinfo {author} {\bibfnamefont {J.~J.~L.}\ \bibnamefont
  {Morton}}, \bibinfo {author} {\bibfnamefont {S.~C.}\ \bibnamefont
  {Benjamin}}, \bibinfo {author} {\bibfnamefont {A.}~\bibnamefont {Ardavan}},
  \bibinfo {author} {\bibfnamefont {G.~A.~D.}\ \bibnamefont {Briggs}}, \bibinfo
  {author} {\bibfnamefont {J.~W.}\ \bibnamefont {Ager}},\ and\ \bibinfo
  {author} {\bibfnamefont {S.~A.}\ \bibnamefont {Lyon}},\ }\href
  {https://doi.org/10.1088/0953-8984/18/21/S06} {\bibfield  {journal} {\bibinfo
   {journal} {Journal of Physics-Condensed Matter}\ }\textbf {\bibinfo {volume}
  {18}},\ \bibinfo {pages} {S783} (\bibinfo {year} {2006})}\BibitemShut
  {NoStop}%
\bibitem [{\citenamefont {Gritsch}\ \emph {et~al.}(2023)\citenamefont
  {Gritsch}, \citenamefont {Ulanowski},\ and\ \citenamefont
  {Reiserer}}]{Gritsch2023}%
  \BibitemOpen
  \bibfield  {author} {\bibinfo {author} {\bibfnamefont {A.}~\bibnamefont
  {Gritsch}}, \bibinfo {author} {\bibfnamefont {A.}~\bibnamefont {Ulanowski}},\
  and\ \bibinfo {author} {\bibfnamefont {A.}~\bibnamefont {Reiserer}},\ }\href
  {https://doi.org/10.1364/OPTICA.486167} {\bibfield  {journal} {\bibinfo
  {journal} {Optica}\ }\textbf {\bibinfo {volume} {10}},\ \bibinfo {pages}
  {783} (\bibinfo {year} {2023})}\BibitemShut {NoStop}%
\bibitem [{\citenamefont {Holmes}\ \emph {et~al.}(2021)\citenamefont {Holmes},
  \citenamefont {Johnson}, \citenamefont {Chua}, \citenamefont {Voisin},
  \citenamefont {Kocsis}, \citenamefont {Rubanov}, \citenamefont {Robson},
  \citenamefont {McCallum}, \citenamefont {McCamey}, \citenamefont {Rogge},\
  and\ \citenamefont {Jamieson}}]{Holmes2021}%
  \BibitemOpen
  \bibfield  {author} {\bibinfo {author} {\bibfnamefont {D.}~\bibnamefont
  {Holmes}}, \bibinfo {author} {\bibfnamefont {B.~C.}\ \bibnamefont {Johnson}},
  \bibinfo {author} {\bibfnamefont {C.}~\bibnamefont {Chua}}, \bibinfo {author}
  {\bibfnamefont {B.}~\bibnamefont {Voisin}}, \bibinfo {author} {\bibfnamefont
  {S.}~\bibnamefont {Kocsis}}, \bibinfo {author} {\bibfnamefont
  {S.}~\bibnamefont {Rubanov}}, \bibinfo {author} {\bibfnamefont {S.~G.}\
  \bibnamefont {Robson}}, \bibinfo {author} {\bibfnamefont {J.~C.}\
  \bibnamefont {McCallum}}, \bibinfo {author} {\bibfnamefont {D.~R.}\
  \bibnamefont {McCamey}}, \bibinfo {author} {\bibfnamefont {S.}~\bibnamefont
  {Rogge}},\ and\ \bibinfo {author} {\bibfnamefont {D.~N.}\ \bibnamefont
  {Jamieson}},\ }\href {https://doi.org/10.1103/PhysRevMaterials.5.014601}
  {\bibfield  {journal} {\bibinfo  {journal} {Physical Review Materials}\
  }\textbf {\bibinfo {volume} {5}},\ \bibinfo {pages} {014601} (\bibinfo {year}
  {2021})}\BibitemShut {NoStop}%
\bibitem [{\citenamefont {Liu}\ \emph {et~al.}(2022)\citenamefont {Liu},
  \citenamefont {Rinner}, \citenamefont {Remmele}, \citenamefont {Ernst},
  \citenamefont {Reiserer},\ and\ \citenamefont {Boeck}}]{Liu2022}%
  \BibitemOpen
  \bibfield  {author} {\bibinfo {author} {\bibfnamefont {Y.}~\bibnamefont
  {Liu}}, \bibinfo {author} {\bibfnamefont {S.}~\bibnamefont {Rinner}},
  \bibinfo {author} {\bibfnamefont {T.}~\bibnamefont {Remmele}}, \bibinfo
  {author} {\bibfnamefont {O.}~\bibnamefont {Ernst}}, \bibinfo {author}
  {\bibfnamefont {A.}~\bibnamefont {Reiserer}},\ and\ \bibinfo {author}
  {\bibfnamefont {T.}~\bibnamefont {Boeck}},\ }\href
  {https://doi.org/10.1016/j.jcrysgro.2022.126733} {\bibfield  {journal}
  {\bibinfo  {journal} {Journal of Crystal Growth}\ }\textbf {\bibinfo {volume}
  {593}},\ \bibinfo {pages} {126733} (\bibinfo {year} {2022})}\BibitemShut
  {NoStop}%
\bibitem [{\citenamefont {Sangtawesin}\ \emph {et~al.}(2019)\citenamefont
  {Sangtawesin}, \citenamefont {Dwyer}, \citenamefont {Srinivasan},
  \citenamefont {Allred}, \citenamefont {Rodgers}, \citenamefont {De~Greve},
  \citenamefont {Stacey}, \citenamefont {Dontschuk}, \citenamefont {O'Donnell},
  \citenamefont {Hu}, \citenamefont {Evans}, \citenamefont {Jaye},
  \citenamefont {Fischer}, \citenamefont {Markham}, \citenamefont {Twitchen},
  \citenamefont {Park}, \citenamefont {Lukin},\ and\ \citenamefont
  {de~Leon}}]{Sangtawesin2019}%
  \BibitemOpen
  \bibfield  {author} {\bibinfo {author} {\bibfnamefont {S.}~\bibnamefont
  {Sangtawesin}}, \bibinfo {author} {\bibfnamefont {B.~L.}\ \bibnamefont
  {Dwyer}}, \bibinfo {author} {\bibfnamefont {S.}~\bibnamefont {Srinivasan}},
  \bibinfo {author} {\bibfnamefont {J.~J.}\ \bibnamefont {Allred}}, \bibinfo
  {author} {\bibfnamefont {L.~V.~H.}\ \bibnamefont {Rodgers}}, \bibinfo
  {author} {\bibfnamefont {K.}~\bibnamefont {De~Greve}}, \bibinfo {author}
  {\bibfnamefont {A.}~\bibnamefont {Stacey}}, \bibinfo {author} {\bibfnamefont
  {N.}~\bibnamefont {Dontschuk}}, \bibinfo {author} {\bibfnamefont {K.~M.}\
  \bibnamefont {O'Donnell}}, \bibinfo {author} {\bibfnamefont {D.}~\bibnamefont
  {Hu}}, \bibinfo {author} {\bibfnamefont {D.~A.}\ \bibnamefont {Evans}},
  \bibinfo {author} {\bibfnamefont {C.}~\bibnamefont {Jaye}}, \bibinfo {author}
  {\bibfnamefont {D.~A.}\ \bibnamefont {Fischer}}, \bibinfo {author}
  {\bibfnamefont {M.~L.}\ \bibnamefont {Markham}}, \bibinfo {author}
  {\bibfnamefont {D.~J.}\ \bibnamefont {Twitchen}}, \bibinfo {author}
  {\bibfnamefont {H.}~\bibnamefont {Park}}, \bibinfo {author} {\bibfnamefont
  {M.~D.}\ \bibnamefont {Lukin}},\ and\ \bibinfo {author} {\bibfnamefont
  {N.~P.}\ \bibnamefont {de~Leon}},\ }\href
  {https://doi.org/10.1103/PhysRevX.9.031052} {\bibfield  {journal} {\bibinfo
  {journal} {Physical Review X}\ }\textbf {\bibinfo {volume} {9}},\ \bibinfo
  {pages} {031052} (\bibinfo {year} {2019})}\BibitemShut {NoStop}%
\bibitem [{\citenamefont {MacQuarrie}\ \emph {et~al.}(2021)\citenamefont
  {MacQuarrie}, \citenamefont {Chartrand}, \citenamefont {Higginbottom},
  \citenamefont {Morse}, \citenamefont {Karasyuk}, \citenamefont {Roorda},\
  and\ \citenamefont {Simmons}}]{Macquarrie2021}%
  \BibitemOpen
  \bibfield  {author} {\bibinfo {author} {\bibfnamefont {E.~R.}\ \bibnamefont
  {MacQuarrie}}, \bibinfo {author} {\bibfnamefont {C.}~\bibnamefont
  {Chartrand}}, \bibinfo {author} {\bibfnamefont {D.~B.}\ \bibnamefont
  {Higginbottom}}, \bibinfo {author} {\bibfnamefont {K.~J.}\ \bibnamefont
  {Morse}}, \bibinfo {author} {\bibfnamefont {V.~A.}\ \bibnamefont {Karasyuk}},
  \bibinfo {author} {\bibfnamefont {S.}~\bibnamefont {Roorda}},\ and\ \bibinfo
  {author} {\bibfnamefont {S.}~\bibnamefont {Simmons}},\ }\href
  {https://doi.org/10.1088/1367-2630/ac291f} {\bibfield  {journal} {\bibinfo
  {journal} {New Journal of Physics}\ }\textbf {\bibinfo {volume} {23}},\
  \bibinfo {pages} {103008} (\bibinfo {year} {2021})}\BibitemShut {NoStop}%
\bibitem [{\citenamefont {Dhaliah}\ \emph {et~al.}(2022)\citenamefont
  {Dhaliah}, \citenamefont {Xiong}, \citenamefont {Sipahigil}, \citenamefont
  {Griffin},\ and\ \citenamefont {Hautier}}]{Dhaliah2022}%
  \BibitemOpen
  \bibfield  {author} {\bibinfo {author} {\bibfnamefont {D.}~\bibnamefont
  {Dhaliah}}, \bibinfo {author} {\bibfnamefont {Y.~H.}\ \bibnamefont {Xiong}},
  \bibinfo {author} {\bibfnamefont {A.}~\bibnamefont {Sipahigil}}, \bibinfo
  {author} {\bibfnamefont {S.~M.}\ \bibnamefont {Griffin}},\ and\ \bibinfo
  {author} {\bibfnamefont {G.}~\bibnamefont {Hautier}},\ }\href
  {https://doi.org/10.1103/PhysRevMaterials.6.L053201} {\bibfield  {journal}
  {\bibinfo  {journal} {Physical Review Materials}\ }\textbf {\bibinfo {volume}
  {6}},\ \bibinfo {pages} {L053201} (\bibinfo {year} {2022})}\BibitemShut
  {NoStop}%
\bibitem [{\citenamefont {DeAbreu}\ \emph {et~al.}(2023)\citenamefont
  {DeAbreu}, \citenamefont {Bowness}, \citenamefont {Alizadeh}, \citenamefont
  {Chartrand}, \citenamefont {Brunelle}, \citenamefont {MacQuarrie},
  \citenamefont {Lee-Hone}, \citenamefont {Ruether}, \citenamefont {Kazemi},
  \citenamefont {Kurkjian}, \citenamefont {Roorda}, \citenamefont {Abrosimov},
  \citenamefont {Pohl}, \citenamefont {Thewalt}, \citenamefont {Higginbottom},\
  and\ \citenamefont {Simmons}}]{DeAbreau2022}%
  \BibitemOpen
  \bibfield  {author} {\bibinfo {author} {\bibfnamefont {A.}~\bibnamefont
  {DeAbreu}}, \bibinfo {author} {\bibfnamefont {C.}~\bibnamefont {Bowness}},
  \bibinfo {author} {\bibfnamefont {A.}~\bibnamefont {Alizadeh}}, \bibinfo
  {author} {\bibfnamefont {C.}~\bibnamefont {Chartrand}}, \bibinfo {author}
  {\bibfnamefont {N.~A.}\ \bibnamefont {Brunelle}}, \bibinfo {author}
  {\bibfnamefont {E.~R.}\ \bibnamefont {MacQuarrie}}, \bibinfo {author}
  {\bibfnamefont {N.~R.}\ \bibnamefont {Lee-Hone}}, \bibinfo {author}
  {\bibfnamefont {M.}~\bibnamefont {Ruether}}, \bibinfo {author} {\bibfnamefont
  {M.}~\bibnamefont {Kazemi}}, \bibinfo {author} {\bibfnamefont {A.~T.~K.}\
  \bibnamefont {Kurkjian}}, \bibinfo {author} {\bibfnamefont {S.}~\bibnamefont
  {Roorda}}, \bibinfo {author} {\bibfnamefont {N.~V.}\ \bibnamefont
  {Abrosimov}}, \bibinfo {author} {\bibfnamefont {H.~J.}\ \bibnamefont {Pohl}},
  \bibinfo {author} {\bibfnamefont {M.~L.~W.}\ \bibnamefont {Thewalt}},
  \bibinfo {author} {\bibfnamefont {D.~B.}\ \bibnamefont {Higginbottom}},\ and\
  \bibinfo {author} {\bibfnamefont {S.}~\bibnamefont {Simmons}},\ }\href
  {https://doi.org/10.1364/Oe.482008} {\bibfield  {journal} {\bibinfo
  {journal} {Optics Express}\ }\textbf {\bibinfo {volume} {31}},\ \bibinfo
  {pages} {15045} (\bibinfo {year} {2023})}\BibitemShut {NoStop}%
\bibitem [{\citenamefont {Sun}(2005)}]{Liu2005}%
  \BibitemOpen
  \bibfield  {author} {\bibinfo {author} {\bibfnamefont {Y.~C.}\ \bibnamefont
  {Sun}},\ }in\ \href {https://doi.org/10.1007/3-540-28209-2} {\emph {\bibinfo
  {booktitle} {Spectroscopic Properties of Rare Earths in Optical
  Materials}}},\ \bibinfo {series and number} {Springer series in materials
  science},\ \bibinfo {editor} {edited by\ \bibinfo {editor} {\bibfnamefont
  {G.}~\bibnamefont {Liu}}\ and\ \bibinfo {editor} {\bibfnamefont
  {B.}~\bibnamefont {Jacquier}}}\ (\bibinfo  {publisher} {Springer Berlin,
  Heidelberg},\ \bibinfo {address} {Berlin},\ \bibinfo {year} {2005})\ p.\
  \bibinfo {pages} {412}\BibitemShut {NoStop}%
\bibitem [{\citenamefont {Kobayashi}\ \emph {et~al.}(2021)\citenamefont
  {Kobayashi}, \citenamefont {Salfi}, \citenamefont {Chua}, \citenamefont
  {van~der Heijden}, \citenamefont {House}, \citenamefont {Culcer},
  \citenamefont {Hutchison}, \citenamefont {Johnson}, \citenamefont {McCallum},
  \citenamefont {Riemann}, \citenamefont {Abrosimov}, \citenamefont {Becker},
  \citenamefont {Pohl}, \citenamefont {Simmons},\ and\ \citenamefont
  {Rogge}}]{Kobayashi2021}%
  \BibitemOpen
  \bibfield  {author} {\bibinfo {author} {\bibfnamefont {T.}~\bibnamefont
  {Kobayashi}}, \bibinfo {author} {\bibfnamefont {J.}~\bibnamefont {Salfi}},
  \bibinfo {author} {\bibfnamefont {C.}~\bibnamefont {Chua}}, \bibinfo {author}
  {\bibfnamefont {J.}~\bibnamefont {van~der Heijden}}, \bibinfo {author}
  {\bibfnamefont {M.~G.}\ \bibnamefont {House}}, \bibinfo {author}
  {\bibfnamefont {D.}~\bibnamefont {Culcer}}, \bibinfo {author} {\bibfnamefont
  {W.~D.}\ \bibnamefont {Hutchison}}, \bibinfo {author} {\bibfnamefont {B.~C.}\
  \bibnamefont {Johnson}}, \bibinfo {author} {\bibfnamefont {J.~C.}\
  \bibnamefont {McCallum}}, \bibinfo {author} {\bibfnamefont {H.}~\bibnamefont
  {Riemann}}, \bibinfo {author} {\bibfnamefont {N.~V.}\ \bibnamefont
  {Abrosimov}}, \bibinfo {author} {\bibfnamefont {P.}~\bibnamefont {Becker}},
  \bibinfo {author} {\bibfnamefont {H.~J.}\ \bibnamefont {Pohl}}, \bibinfo
  {author} {\bibfnamefont {M.~Y.}\ \bibnamefont {Simmons}},\ and\ \bibinfo
  {author} {\bibfnamefont {S.}~\bibnamefont {Rogge}},\ }\href
  {https://doi.org/10.1038/s41563-020-0743-3} {\bibfield  {journal} {\bibinfo
  {journal} {Nature Materials}\ }\textbf {\bibinfo {volume} {20}},\ \bibinfo
  {pages} {38} (\bibinfo {year} {2021})}\BibitemShut {NoStop}%
\bibitem [{\citenamefont {Yin}\ \emph {et~al.}(2013)\citenamefont {Yin},
  \citenamefont {Rancic}, \citenamefont {de~Boo}, \citenamefont {Stavrias},
  \citenamefont {McCallum}, \citenamefont {Sellars},\ and\ \citenamefont
  {Rogge}}]{Yin2013}%
  \BibitemOpen
  \bibfield  {author} {\bibinfo {author} {\bibfnamefont {C.~M.}\ \bibnamefont
  {Yin}}, \bibinfo {author} {\bibfnamefont {M.}~\bibnamefont {Rancic}},
  \bibinfo {author} {\bibfnamefont {G.~G.}\ \bibnamefont {de~Boo}}, \bibinfo
  {author} {\bibfnamefont {N.}~\bibnamefont {Stavrias}}, \bibinfo {author}
  {\bibfnamefont {J.~C.}\ \bibnamefont {McCallum}}, \bibinfo {author}
  {\bibfnamefont {M.~J.}\ \bibnamefont {Sellars}},\ and\ \bibinfo {author}
  {\bibfnamefont {S.}~\bibnamefont {Rogge}},\ }\href
  {https://doi.org/10.1038/nature12081} {\bibfield  {journal} {\bibinfo
  {journal} {Nature}\ }\textbf {\bibinfo {volume} {497}},\ \bibinfo {pages}
  {91} (\bibinfo {year} {2013})}\BibitemShut {NoStop}%
\bibitem [{\citenamefont {Palm}\ \emph {et~al.}(1996)\citenamefont {Palm},
  \citenamefont {Gan}, \citenamefont {Zheng}, \citenamefont {Michel},\ and\
  \citenamefont {Kimerling}}]{Palm1996}%
  \BibitemOpen
  \bibfield  {author} {\bibinfo {author} {\bibfnamefont {J.}~\bibnamefont
  {Palm}}, \bibinfo {author} {\bibfnamefont {F.}~\bibnamefont {Gan}}, \bibinfo
  {author} {\bibfnamefont {B.}~\bibnamefont {Zheng}}, \bibinfo {author}
  {\bibfnamefont {J.}~\bibnamefont {Michel}},\ and\ \bibinfo {author}
  {\bibfnamefont {L.~C.}\ \bibnamefont {Kimerling}},\ }\href
  {https://doi.org/10.1103/PhysRevB.54.17603} {\bibfield  {journal} {\bibinfo
  {journal} {Physical Review B}\ }\textbf {\bibinfo {volume} {54}},\ \bibinfo
  {pages} {17603} (\bibinfo {year} {1996})}\BibitemShut {NoStop}%
\bibitem [{\citenamefont {Taguchi}\ \emph {et~al.}(1998)\citenamefont
  {Taguchi}, \citenamefont {Takahei}, \citenamefont {Matsuoka},\ and\
  \citenamefont {Tohno}}]{Taguchi1998}%
  \BibitemOpen
  \bibfield  {author} {\bibinfo {author} {\bibfnamefont {A.}~\bibnamefont
  {Taguchi}}, \bibinfo {author} {\bibfnamefont {K.}~\bibnamefont {Takahei}},
  \bibinfo {author} {\bibfnamefont {M.}~\bibnamefont {Matsuoka}},\ and\
  \bibinfo {author} {\bibfnamefont {S.}~\bibnamefont {Tohno}},\ }\href
  {https://doi.org/10.1063/1.368673} {\bibfield  {journal} {\bibinfo  {journal}
  {Journal of Applied Physics}\ }\textbf {\bibinfo {volume} {84}},\ \bibinfo
  {pages} {4471} (\bibinfo {year} {1998})}\BibitemShut {NoStop}%
\bibitem [{\citenamefont {Priolo}\ \emph {et~al.}(1998)\citenamefont {Priolo},
  \citenamefont {Franzo}, \citenamefont {Coffa},\ and\ \citenamefont
  {Carnera}}]{Priolo1998}%
  \BibitemOpen
  \bibfield  {author} {\bibinfo {author} {\bibfnamefont {F.}~\bibnamefont
  {Priolo}}, \bibinfo {author} {\bibfnamefont {G.}~\bibnamefont {Franzo}},
  \bibinfo {author} {\bibfnamefont {S.}~\bibnamefont {Coffa}},\ and\ \bibinfo
  {author} {\bibfnamefont {A.}~\bibnamefont {Carnera}},\ }\href
  {https://doi.org/10.1103/PhysRevB.57.4443} {\bibfield  {journal} {\bibinfo
  {journal} {Physical Review B}\ }\textbf {\bibinfo {volume} {57}},\ \bibinfo
  {pages} {4443} (\bibinfo {year} {1998})}\BibitemShut {NoStop}%
\bibitem [{\citenamefont {Car}\ \emph {et~al.}(2020)\citenamefont {Car},
  \citenamefont {Le~Gouet},\ and\ \citenamefont {Chaneliere}}]{Car2020}%
  \BibitemOpen
  \bibfield  {author} {\bibinfo {author} {\bibfnamefont {B.}~\bibnamefont
  {Car}}, \bibinfo {author} {\bibfnamefont {J.~L.}\ \bibnamefont {Le~Gouet}},\
  and\ \bibinfo {author} {\bibfnamefont {T.}~\bibnamefont {Chaneliere}},\
  }\href {https://doi.org/10.1103/PhysRevB.102.115119} {\bibfield  {journal}
  {\bibinfo  {journal} {Physical Review B}\ }\textbf {\bibinfo {volume}
  {102}},\ \bibinfo {pages} {115119} (\bibinfo {year} {2020})}\BibitemShut
  {NoStop}%
\bibitem [{\citenamefont {Probst}\ \emph {et~al.}(2020)\citenamefont {Probst},
  \citenamefont {Zhang}, \citenamefont {Ran\v{c}i\'c}, \citenamefont {Ranjan},
  \citenamefont {Le~Dantec}, \citenamefont {Zhang}, \citenamefont {Albanese},
  \citenamefont {Doll}, \citenamefont {Liu}, \citenamefont {Morton},
  \citenamefont {Chaneli\`ere}, \citenamefont {Goldner}, \citenamefont {Vion},
  \citenamefont {Esteve},\ and\ \citenamefont {Bertet}}]{Probst2020}%
  \BibitemOpen
  \bibfield  {author} {\bibinfo {author} {\bibfnamefont {S.}~\bibnamefont
  {Probst}}, \bibinfo {author} {\bibfnamefont {G.}~\bibnamefont {Zhang}},
  \bibinfo {author} {\bibfnamefont {M.}~\bibnamefont {Ran\v{c}i\'c}}, \bibinfo
  {author} {\bibfnamefont {V.}~\bibnamefont {Ranjan}}, \bibinfo {author}
  {\bibfnamefont {M.}~\bibnamefont {Le~Dantec}}, \bibinfo {author}
  {\bibfnamefont {Z.}~\bibnamefont {Zhang}}, \bibinfo {author} {\bibfnamefont
  {B.}~\bibnamefont {Albanese}}, \bibinfo {author} {\bibfnamefont
  {A.}~\bibnamefont {Doll}}, \bibinfo {author} {\bibfnamefont {R.~B.}\
  \bibnamefont {Liu}}, \bibinfo {author} {\bibfnamefont {J.}~\bibnamefont
  {Morton}}, \bibinfo {author} {\bibfnamefont {T.}~\bibnamefont
  {Chaneli\`ere}}, \bibinfo {author} {\bibfnamefont {P.}~\bibnamefont
  {Goldner}}, \bibinfo {author} {\bibfnamefont {D.}~\bibnamefont {Vion}},
  \bibinfo {author} {\bibfnamefont {D.}~\bibnamefont {Esteve}},\ and\ \bibinfo
  {author} {\bibfnamefont {P.}~\bibnamefont {Bertet}},\ }\href
  {https://doi.org/10.5194/mr-1-315-2020} {\bibfield  {journal} {\bibinfo
  {journal} {Magnetic Resonance}\ }\textbf {\bibinfo {volume} {1}},\ \bibinfo
  {pages} {315} (\bibinfo {year} {2020})}\BibitemShut {NoStop}%
\end{thebibliography}%

\end{document}